\documentclass[reprint,
superscriptaddress,
showpacs,
nofootinbib,
prl,
aps,
floatfix,
]{revtex4-1}

\usepackage{amsmath,amssymb,amsfonts,amsthm}
\usepackage[english]{babel}
\usepackage{url}
\usepackage{bm}
\usepackage[latin1]{inputenc}
\usepackage{graphicx,rotating}
\usepackage[colorlinks=true,linkcolor=black, citecolor=blue]{hyperref}

\usepackage{mathtools}
\newcommand{\slashed}[1]{\ensuremath{\mathrlap{\!\not{\phantom{#1}}}#1}}

\begin{document}
\title{Radiative corrections to nucleon weak charges and Beyond Standard Model impact}
\author{Leendert Hayen}
\email[]{lmhayen@ncsu.edu}
\affiliation{Department of Physics, North Carolina State University, Raleigh, North Carolina 27695, USA}
\affiliation{Triangle Universities Nuclear Laboratory, Durham, North Carolina 27708, USA}
\affiliation{Instituut voor Kern- en Stralingsfysica, KU Leuven, Celestijnenlaan 200D, B-3001 Leuven, Belgium}

\date{\today}
\begin{abstract}
The nucleon axial charge is a central ingredient in nuclear and particle physics, and a key observable in precision tests of the electroweak Standard Model sector and beyond. We report on the first complete calculation of its electroweak quantum corrections up to $\mathcal{O}(\alpha)$, using a combination of current algebra techniques and QCD sum rules. We find a substantial enhancement due to the weak magnetism contribution in the elastic channel, and include higher-twist and target mass corrections at low $Q^2$ to find $\Delta_R^A = 0.02881(22)$. Using analogous methods, we determine a new value for the vector charge renormalization, $\Delta_R^V = 0.02474(27)$, and show how the two most recent calculations can be brought into agreement. This allows us to determine a corrected experimental $g_A^0 = 1.2730(13)$, which is a $>2\sigma$ shift away from the commonly quoted value. We use this new result to set constraints on exotic right-handed currents by comparing to lattice QCD results, and resolve a double-counting issue in the $|V_{ud}|$ extraction from mirror decays.
\end{abstract}

\maketitle

Precision studies of neutron $\beta$ decay put stringent tests on extensions of the charged weak sector at the TeV scale with precision equal to or exceeding that of colliders \cite{Gonzalez-Alonso2018, Wauters2014, Vos2015, Cirigliano2013, Cirigliano2013b, Pattie2013}. In particular, studies of the unitarity requirement of the Cabibbo-Kobayashi-Maskawa (CKM) matrix and comparison to lattice QCD calculations have shown tremendous reach in several Beyond Standard Model (BSM) channels \cite{Cirgiliano2019, Alioli2017}. This is possible only through exquisite control of strong and electroweak quantum corrections to the tree-level vector-axial vector ($V$-$A$) weak interaction \cite{Sirlin2013, Czarnecki2018}.

Even though the vector current is protected from the strong interaction \cite{Ademollo1964}, QCD effects shift the nucleon axial-vector coupling constant, $g_A$, away from unity. In typical charged current processes such as (nuclear) $\beta$ decay, however, these bare quantities are not observable experimentally as both $g_V$ and $g_A$ contain percent-level corrections due to electroweak radiative processes
\begin{equation}
    g_L^2 \to g_L^2(1+\Delta_R^L)
\end{equation}
with $L=V, A$. Up to now, however, only $\Delta_R^V$ has received much attention due to its importance in the $|V_{ud}|$ extraction from pure vector transitions, with $V_{ud}$ the up-down CKM matrix element \cite{Sirlin1978, Seng2018, Seng2019, Czarnecki2019}. Traditionally, for mixed transitions such as the neutron one ignores this difference and defines the neutron lifetime, $\tau_n$, by extracting the vector radiative corrections as follows
\begin{equation}
    (1+\Delta_R^V)\left[g_V^2f_V+3(g_A^\mathrm{eff})^2f_A\right] = \frac{K}{|V_{ud}|^2G_F^2}\frac{1}{\tau_n}
    \label{eq:general_tiple_relation}
\end{equation}
with $K$ a collection of constants, $G_F$ the Fermi coupling constant and $f_{V, A}$ so-called phase space factors. Equation \ref{eq:general_tiple_relation} implies an `operational' definition of $g_A$ \cite{Czarnecki2018}
\begin{equation}
    g_A^\mathrm{eff} = g_A^\text{QCD}\left[1 + \left(\Delta_R^A - \Delta_R^V\right)/2\right]\left[1 - 2\, \text{Re}~\epsilon_R \right],
    \label{eq:gA_def_QCD}
\end{equation}
where we added a possible contamination from a BSM right-handed coupling, $\epsilon_R \propto M_W^2/\Lambda_{BSM}^2$ \cite{Bhattacharya2012}. This definition can be made in the tree-level Lagrangian so that all experimental observables probe $g_A^\mathrm{eff}$ instead. With the rise of precision lattice QCD (LQCD) calculations for $g_A^\mathrm{QCD}$ \cite{Gupta2018, Chang2018, Aoki2020}, comparison with experimental $g_A^\mathrm{eff}$ results was recognized as an extremely clean way of probing $\epsilon_R$ \cite{Alioli2017}. A correct assessment, however, relies on a clean separation of electroweak and strong effects.

In this Letter, we present a first complete Standard Model calculation of electroweak radiative corrections (EWRC) to the neutron axial vector charge, $g_A$, and a new result for $\Delta_R^V$. Using a combination of current algebra techniques with QCD sum rules, we find
\begin{subequations}
\begin{align}
    \Delta_R^A &= 0.02881(22), \label{eq:Delta_R_A_final_intro} \\
    \Delta_R^V &= 0.02474(27) \label{eq:Delta_R_V_final_intro}
\end{align}
\end{subequations}
and show how the two most recent $\Delta_R^V$ results can be made in agreement \cite{Seng2019, Czarnecki2019}. As a consequence, we can for the first time extract
\begin{equation}
    g_A^{0} \equiv \frac{g_A^\mathrm{eff}}{1+(\Delta_R^A-\Delta_R^V)/2} = 1.2730(13),
\end{equation}
which is a $>2\sigma$ departure from its quoted value, $g_A^\mathrm{eff} = 1.2756(13)$ \cite{Zyla2020}, and which is the value to be used in neutral current processes and compared to LQCD. Additionally, we resolve a double-counting issue in the $|V_{ud}|$ extraction from $T=1/2$ mirror systems, and provide a new $|V_{ud}|$ value from superallowed $0^+\to 0^+$ decays \cite{Hardy2020} from Eq. (\ref{eq:Delta_R_V_final_intro}), 
\begin{subequations}
\begin{align}
    |V_{ud}|^\mathrm{mirror} &= 0.9730(10), \\
    |V_{ud}|^\mathrm{super} &= 0.9736(3).
\end{align}
\end{subequations}
We summarize in this Letter the essential features of our analysis, and defer details to a longer paper \cite{Hayen2020c}.

\begin{figure}[t]
    \centering
    \includegraphics[width=0.2\textwidth]{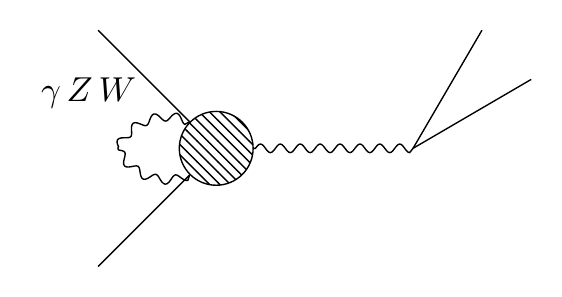}
    \includegraphics[width=0.2\textwidth]{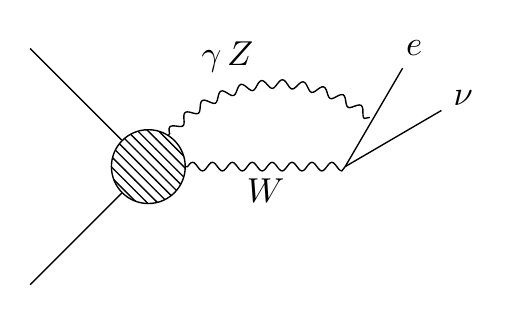}
    \caption{\label{fig:feynman_order_alpha}$\mathcal{O}(\alpha)$ radiative corrections that give rise to differences in vector and axial vector transitions.}
\end{figure}

The small momentum exchange in $\beta$ decay ($q \sim$ MeV) relative to the $W$-boson mass limits the number of diagrams that contribute to order $\mathcal{O}(G_F\alpha)$ with $\alpha$ the fine-structure constant, summarized in the seminal work by Sirlin \cite{Sirlin1978}. Further, when evaluating Eq. (\ref{eq:gA_def_QCD}) we need to worry only about those diagrams which depend on the vector or axial-vector nature of the transition. Differences are anticipated \textit{a priori} in the vertex and box correction diagrams, shown in Fig. \ref{fig:feynman_order_alpha}. It is straightforward, however, to show that only diagrams containing virtual photons contribute to first order, as those with either $Z$ or $W$ bosons are infrared convergent and are as such either higher-order or common to both $\Delta_R^L$.

The total EWRC can be parametrized according to \cite{Czarnecki2004, Czarnecki2019}
\begin{equation}
    \Delta_R^{L} = 0.01686 + 1.022L_{NP} + 1.065L_P
    \label{eq:Delta_R_I_param}
\end{equation}
where $(N)P$ designates (non-)perturbative contributions, respectively. The first term combines large-log terms originating from high-momentum behaviour common to vector and axial transitions \cite{Sirlin1982}, and includes the asymptotic $ZW$ box contribution \cite{Czarnecki2019}.

Starting with the photonic vertex correction, we can use the on mass-shell renormalization theorem which states that the modification of a vertex function $\langle p_f | \Gamma^\mu | p_i \rangle = F^\mu(p_f, p_i)$ due to an interaction term $\delta \mathcal{L}$ in the Lagrangian is \cite{Sirlin1978, Seng2019}
\begin{equation}
    \delta F^\mu(p_f, p_i) = \lim_{\bar{q}\to q} i T^\mu(\bar{q}, p_i, p_f),
\end{equation}
where $q=p_i-p_f$ and the tensor $T^{\mu} = T^{\mu\lambda}{}_\lambda(\bar{q}, p_f, p_i)$ is
\begin{equation}
    T^{\mu\lambda}{}_\lambda = \int d^4x\, e^{i\bar{q}x}\langle p_f | T\{J^\mu_W(x) \delta \mathcal{L}(0)\} | p_i \rangle - B^\mu,
\end{equation}
where $J^\mu_W$ is the weak current and the final term subtracts the poles from the mass renormalization to make $T^\mu$ pole-free by construction. In Fig. \ref{fig:feynman_order_alpha} the interaction is
\begin{equation}
    \delta \mathcal{L}(0) = \frac{\alpha}{(2\pi)^3}\int \frac{d^4k}{k^2}\int d^4y\,e^{iky}~ T\{J^\lambda_\gamma(y)J_\lambda^\gamma(0)\},
\end{equation}
where $J^\gamma_\lambda$ is the photonic current. Using the identity
\begin{equation}
    iT^\mu = -\bar{q}_\nu \frac{\partial}{\partial \bar{q}_\mu}iT^\nu + \frac{\partial}{\partial \bar{q}_\mu}(i\bar{q}_\nu T^\nu),
    \label{eq:T_mu_identity}
\end{equation}
a partial integration of $\bar{q}_\nu T^\nu$ unlocks equal-time current commutators through
\begin{align}
    \frac{\partial}{\partial x^\nu} &T\left\{ J^\nu_W(x)J^\lambda_\gamma(y)J_\lambda^\gamma(0) \right\} = \nonumber \\ &T\biggl\{\partial_\nu J_W^\nu(x) J^\lambda_\gamma(y)J_\lambda^\gamma(0) + \delta(x^0) \left[J^0_W(x), J^\lambda_\gamma(0) \right]J_\lambda^\gamma(y) \nonumber \\
    &+ \delta(x^0-y^0)\left[J^0_W(x), J^\lambda_\gamma(y)\right]J_\lambda^\gamma(0) \biggr\}.
    \label{eq:derivative_three_current}
\end{align}
The commutators can be evaluated immediately using the current algebra relationship
\begin{equation}
    \left[J^0_W(t, \bm{x}), J^\mu_\gamma(t, \bm{y}) \right] =  J^\mu_W(x)\, \delta^{(3)}(\bm{x}-\bm{y}), \label{eq:ETCR_gamma_W}
\end{equation}
which is conserved even in the presence of strong interaction effects \cite{Adler1968, Treiman1972}. From Eq. (\ref{eq:derivative_three_current}) we obtain a three-current, denoted $\mathcal{D}_\gamma$, and two-current piece. The latter can be shown to cancel with an opposite contribution from the $\gamma W$ box diagram and we continue with the former. Since it depends on the divergence of the weak current, the vector part of which is conserved, it vanishes for vector transitions and we anticipate a non-zero contribution for an axial transition. 

In the high-momentum region ($k^2 \gg 1$ GeV$^2$) the divergence vanishes also for an axial transition due to chiral symmetry - or equivalently using the partially conserved axial current. This can also be checked explicitly \cite{Hayen2020c} through, e.g., an operator product expansion or, in this case equivalently, a Bjorken-Johnson-Low limit \cite{Bjorken1966, Johnson1961, Sirlin1968}.

In the low energy limit we can describe the photon and weak coupling for on-shell nucleons, $N$, in the isospin formalism
\begin{subequations}
\begin{align}
    J^{\gamma, I}_\mu &= i \bar{N} \left[F_1^I \gamma_\mu + i \frac{F_2^I}{2M}\sigma_{\mu \nu} q^\nu \right]\tau^{I} N, \label{eq:photon_coupling_Born}\\
    J^W_\mu &= i\bar{N} \left[g_V \gamma_\mu + i \frac{g_M}{2M}\sigma_{\mu\nu}q^\nu +g_A \gamma_\mu\gamma^5 \right] \tau^{\pm}N \label{eq:weak_coupling_Born}
\end{align}
\end{subequations}
with $I=0, 1$ denoting isoscalar and isovector terms, respectively, $F_{1, 2}(q^2)$ are the charge and magnetic form factors, $g_i(q^2)$ are weak form factors, $g_M(0) = \kappa_p-\kappa_n = 3.706$ is the so-called weak magnetism contribution and $\tau$ are the $SU(2)$ generators. By enforcing $G$-parity \cite{Weinberg1958}, the photon currents in $\mathcal{D}_\gamma$ are restricted to be either both isoscalar or both isovector. Using the fact that $\partial^\mu J_\mu^W$ must be proportional to $\tau^\pm$ it follows after some algebra that
\begin{align}
    D_\gamma^\mathrm{Born} &\approx \biggl[(F_1^0)^2-(F_1^1)^2\biggr]2g_AM [\bar{N}' \gamma^5 \tau^{\pm} N] \nonumber \\
    &\times \int \frac{d^4k}{k^2} \frac{\Lambda^2}{\Lambda^2-k^2} \frac{1}{k_0^2 + i\epsilon}.
    \label{eq:D_gamma_cancellation_scalar_vector}
\end{align}
where $\Lambda$ is a Pauli-Villars mass regulator. In the isospin limit and for small momenta we have $F_1^0(0) = F_1^1(0) = 1$ and the contribution vanishes also at low energy.

Moving on, the matrix element for the photonic box diagram in Fig. \ref{fig:feynman_order_alpha} can be written down as \cite{Sirlin1978}
\begin{equation}
    \mathcal{M}_{\gamma W} = \frac{\alpha G_FV_{ud}}{2^{3/2}\pi^3} \int \frac{d^4k}{k^2} \frac{\bar{e}(2l^\mu - \gamma^\mu\slashed{k})\gamma^\nu (1-\gamma^5)\nu}{[k^2 - 2l\cdot k][1-k^2/M_W^2]} T_{\mu \nu}^{\gamma W}
    \label{eq:gamma_W_box_def}
\end{equation}
where $k$ and $l$ are virtual loop and outgoing lepton four-momenta, respectively, and $T_{\mu\nu}^{\gamma W}$ describes the blob in Fig. \ref{fig:feynman_order_alpha}
\begin{equation}
    T_{\mu\nu}^\gamma(k) = \int d^4x e^{ikx} \langle p_f | T\{J_\mu^\gamma(x)J_\nu^W(0)\} | p_i \rangle. \label{eq:T_gamma}
\end{equation}
Equation (\ref{eq:gamma_W_box_def}) can be reduced by expanding the $\gamma$ matrix product and using Ward-Takahashi identities
\begin{subequations}
\begin{align}
    k^\mu T_{\mu\nu}^{\gamma W} &= i \langle p_f | J_\nu^W | p_i \rangle
    \label{eq:WTI_qed} \\
    k^\nu T_{\mu\nu}^{\gamma W} &= i \langle p_f | J_\mu^W | p_i \rangle + q^\nu T_{\mu \nu} \nonumber \\
    &+ i\int d^4x e^{i (k - q) \cdot x} \langle p_f | T\{\partial^\nu J_\nu^W J_\mu^{\gamma} \} | p_i \rangle.
    \label{eq:WTI_weak}
\end{align}
\end{subequations}
where we used the conservation of the QED current. Terms proportional to the tree-level interaction drop out in Eq. (\ref{eq:gA_def_QCD}), and after neglecting terms of $\mathcal{O}(\alpha q)$ and combining other finite terms with other diagrams, contributions arise from two terms
\begin{equation}
    \mathcal{M}_{\gamma W} =  \frac{\alpha G_FV_{ud}}{4\pi^3}L_\alpha \int \frac{d^4k}{k^4} \frac{[i \epsilon^{\mu \lambda \nu \alpha} k_\lambda T_{\mu \nu}^{\gamma W} - \mathcal{D}^\alpha]}{1-k^2/M_W^2}
    \label{eq:gamma_W_box_WTI}
\end{equation}
where we have taken $l, m_e \to 0$ since both are IR convergent. The first term is the parity-violating contribution, resulting in the infamous axial-vector contribution to a vector transition and vice versa. The second, $\mathcal{D}^\alpha$, corresponds to the second line in Eq. (\ref{eq:WTI_weak}) and depends on the divergence of the weak current responsible for the transition. This resembles the three-current correlator discussed above and so in principle introduces a difference in $\Delta_R^L$. Analogous to our discussion of the vertex correction above, however, the high-energy behaviour of $\mathcal{D}^\alpha$ can be shown to vanish due to chiral symmetry.

The low-energy behaviour, on the other hand, can be treated similarly to Eq. (\ref{eq:D_gamma_cancellation_scalar_vector}) by once more invoking $G$-parity meaning only the isovector photons contribute and we find
\begin{align}
    \mathcal{D}_\mu^\mathrm{Born} \approx -i\pi\delta(k_0) F_1^1 \bar{N} \left[\tau^z \partial_\nu J^\nu_W + \partial_\nu J^\nu_W \tau^z \right] N,
\end{align}
neglecting the $\mathcal{O}(1/M)$ magnetic term. Since $\partial^\nu J_\nu^W \propto \tau^{\pm}$ the contribution vanishes by crossing symmetry.

This leaves the analogous contribution of the $\gamma W$ box to axial transitions as the one that has occupied research for vector transitions over the past half-century. Recent work \cite{Seng2018, Seng2019, Czarnecki2019} has shown that contributions from physics at intermediate scales are significant and extend into the elastic regime. Going forward, we simplify the notation of the remaining term in the box contribution of Eq. (\ref{eq:gamma_W_box_WTI}) by introducing a general function $F^{A,V}(Q^2)$ \cite{Marciano2006}
\begin{equation}
    \Box_{\gamma W} = \frac{\alpha}{2\pi} \int_0^\infty dQ^2\frac{M_W^2}{Q^2+M_W^2} F^{V,A}(Q^2)
    \label{eq:box_Q2_general}
\end{equation} 
where $Q^2 = -k^2$ is the virtuality. The Born contribution can be written down easily using Eqs. (\ref{eq:photon_coupling_Born})-(\ref{eq:weak_coupling_Born}) and some algebra results in
\begin{subequations}
\begin{align}
    F^V_\mathrm{Born} &= \frac{1}{Q^2}g_A (F_1+F_2) P \\
    F^A_\mathrm{Born} &= \frac{1}{Q^2}\biggl[g_V (F_1+F_2)+g_MF_1\biggr]P \label{eq:F_A_Born}
\end{align}
\end{subequations}
neglecting $\mathcal{O}(1/M)$ terms and where $P$ is the kinematic contribution from the nucleon propagators
\begin{equation}
    P = \dfrac{1+2\sqrt{1+4M^2/Q^2}}{[1+\sqrt{1+4M^2/Q^2}]^2}
\end{equation}
We can once more use, e.g., $G$-parity to show that only isoscalar photons contribute, which is a well-known result \cite{Sirlin1967}. We perform the integral in Eq. (\ref{eq:box_Q2_general}) numerically over the full range using experimental data for $g_V(Q^2)$ and $g_M(Q^2)$ \cite{Ye2018} and input from Ref. \cite{Bhattacharya2011} for $g_A(Q^2)$ to find
\begin{subequations}
\begin{align}
    V_{NP} &= 0.91(5)\frac{\alpha}{\pi}\\
    A_{NP} &= \left[0.51(3) + 2.13(3)\right]\frac{\alpha}{\pi}
\end{align}
\end{subequations}
where we explicitly separated contributions due to $g_V$ and $g_M$ in the second line, and assign uncertainties as in Ref. \cite{Seng2019}. Interestingly, the weak magnetism contribution clearly dominates for axial transitions and provides a much larger enhancement than anticipated.

\begin{figure}[t]
    \centering
    \includegraphics[width=0.48\textwidth]{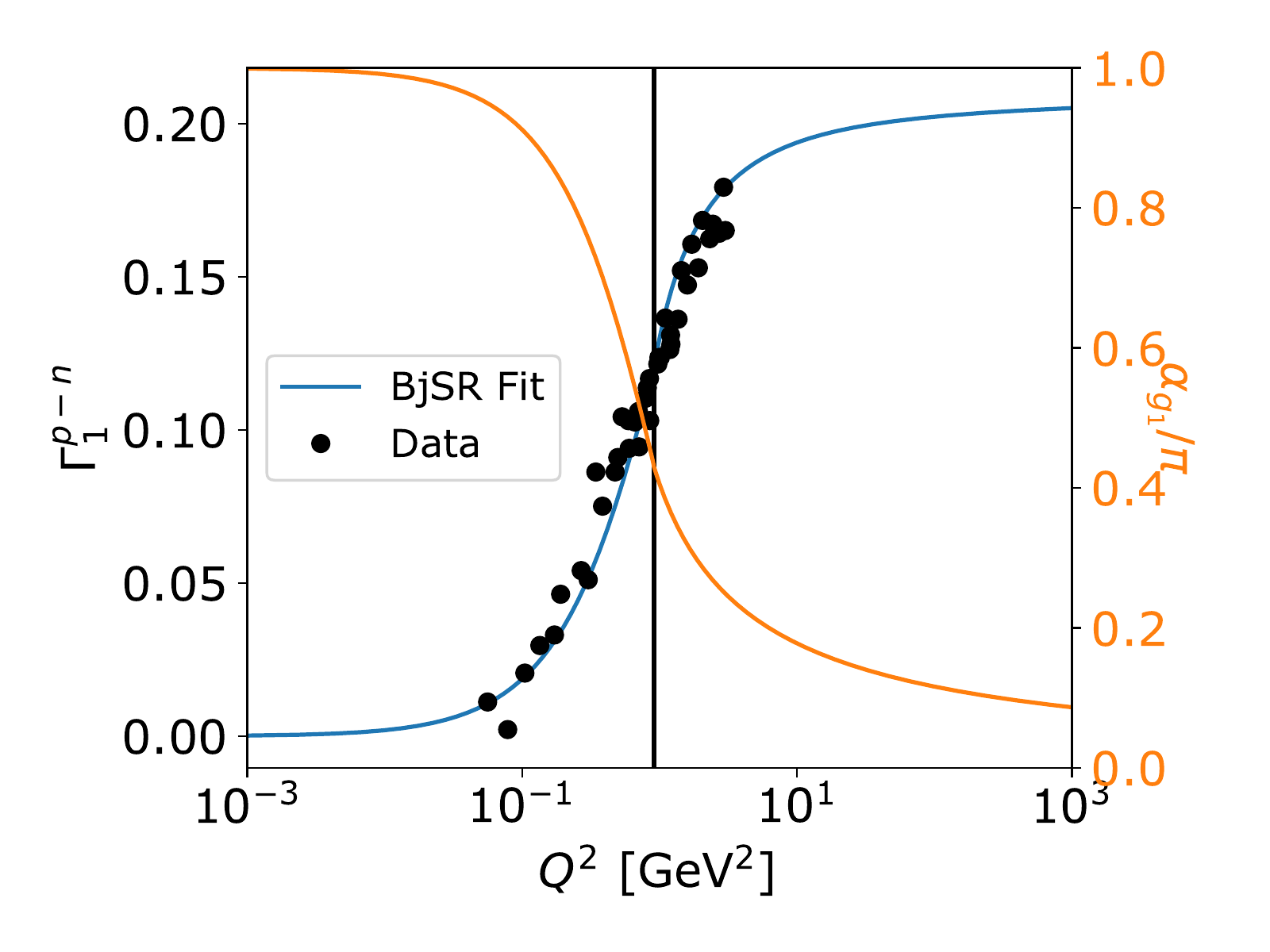}
    \caption{Parametrization of the PBjSR and running coupling $\alpha_{g_1}(Q^2)/\pi$ using the pQCD $\overline{MS}$ parametrization, Eq. (\ref{eq:PBjSR_DIS}), for $Q^2 > Q_0^2 = 0.910$ GeV$^2$ and the LFH result of Eq. (\ref{eq:PBjSR_LFH}) for $Q^2 \leq Q_0^2$, together with experimental data between 0.05 GeV$^2$ and 3 GeV$^2$. Figure adopted from Ref. \cite{Ayala2018a}.}
    \label{fig:PBjSr}
\end{figure}

For the remaining momentum range we take inspiration from the idea originally by Marciano and Sirlin \cite{Marciano2006} and its recent extension \cite{Czarnecki2019}. We may decompose Eq. (\ref{eq:T_gamma}) into structure functions after spin summation
\begin{align}
    &\sum_\mathrm{spins} T^{\mu\nu} \stackrel{\text{asy}}{\longrightarrow} i\epsilon^{\mu\nu\alpha\beta} \frac{k_\alpha p_\beta}{2 (p\cdot k)}F_3(\nu, Q^2) \nonumber \\ 
    &+i\epsilon^{\mu\nu\alpha\beta}\frac{q_\alpha}{p\cdot k}\biggl[S_\beta g_1(\nu, Q^2) + \left(S_\beta-p_\beta\frac{S\cdot k}{p\cdot k}\right) g_2(\nu, Q^2)\biggr],
    \label{eq:expansion_structure_functions_axial_asy}
\end{align}
with $S_\beta$ the four-polarization, $\nu = p\cdot k/M$ and writing only terms that survive the contraction with the Levi-Cevita tensor in Eq. (\ref{eq:gamma_W_box_WTI}). For an axial transition, we may relate Eq. (\ref{eq:T_gamma}) to the Bjorken (Bj) sum rule of polarized electron scattering through an isospin rotation, i.e. $\gamma W \to \gamma \gamma$, specifically
\begin{equation}
    \int_0^{1}dx [g_1^p(x, Q^2)-g_1^n(x, Q^2)] = \frac{g_A}{6}\left[1-\frac{\alpha_{g_1}(Q^2)}{\pi} \right],
    \label{eq:PBjSR_DIS}
\end{equation}
where $x = Q^2/2M\nu$ and $\alpha_{g_1}(Q^2)$ is an effective QCD coupling constant. In the isospin limit, the running of Eq. (\ref{eq:PBjSR_DIS}) is identical to that of $T^{\mu\nu}_{\gamma}$ for axial transitions, and so
\begin{equation}
    F^{A}_\mathrm{DIS}(Q^2) \approx \frac{1}{4Q^2}\left[1 - \frac{\alpha_{g_1}(Q^2)}{\pi}\right].
    \label{eq:F_A_DIS}
\end{equation}
We can take advantage of the substantial amount of experimental data of Eq. (\ref{eq:PBjSR_DIS}) down to $Q^2 \approx 0.1$ GeV$^2$ \cite{Ayala2018a}, thereby also neatly taking into account low energy contributions shown in Fig. \ref{fig:PBjSr}.

Similar to Refs. \cite{Ayala2018a, Czarnecki2019} we stitch together expressions from perturbative QCD (pQCD) at high $Q^2$ \cite{Baikov2010} and a light-front holography (LFH) motivated expression at low $Q^2$ \cite{Brodsky2010}
\begin{equation}
    1 - \frac{\alpha_{g_1}(Q^2)}{\pi} \stackrel{Q^2 < Q^2_0}{=} 
        1-\exp\left(-\frac{Q^2}{4\kappa^2}\right)
        \label{eq:PBjSR_LFH}
\end{equation}
at a variable cross-over point $Q_0^2$ but keep $\kappa$ a free fit parameter. For the pQCD results we use the latest world average $\alpha_s(M_Z^2) = 0.1179 \pm 0.0010$ \cite{Zyla2020} and a 5 loop $\beta$ function calculation from the \texttt{RunDec} package \cite{Herren2018}. 

Analogous to the axial renormalization, we can use isospin symmetry to relate the vector coupling renormalization to parity-violating (anti)neutrino nucleon scattering. The latter obeys the Gross-Llewellyn Smith (GLS) sum rule \cite{Larin1991}
\begin{equation}
    \int_0^1dx [F_3^{\nu p}(x, Q^2) + F_3^{\bar{\nu}p}(x, Q^2)] = 3 \left[1-\frac{\alpha_{F_3}(Q^2)}{\pi}\right],
    \label{eq:GLS_def} 
\end{equation}
leading to a similar relation as Eq. (\ref{eq:F_A_DIS}). Data is much more sparse for these processes, however, resulting in a larger uncertainty \cite{Kim1998}. We note that in Refs. \cite{Marciano2006, Czarnecki2019} one instead related the vector renormalization running to the polarized Bjorken sum rule using a chiral rotation. At low $Q^2$, however, chiral symmetry is broken and the approach of Ref. \cite{Czarnecki2019} is not strictly rigorous. The difference between GLS and Bj sum rules is small, however, and limited to $\mathcal{O}(\alpha^3)$ effects \cite{Baikov2010}. Numerically we obtain
\begin{equation}
\begin{array}{c}
     \delta V_{NP};~ V_P \\
     \delta A_{NP};~ A_P
\end{array} = \frac{\alpha}{\pi} \left\{
    \begin{array}{cc}
        0.199(42); & 2.015(34) \\
        0.173(30); & 2.025(22)
    \end{array}
    \right.
    \label{eq:sum_rule_contributions}
\end{equation}
where the columns denote contributions from below and above $Q^2_0$, respectively. The uncertainty is smallest for the axial transition due to the stronger experimental constraints of the Bj sum rule.

\begin{figure}[b]
    \centering
    \includegraphics[width=0.48\textwidth]{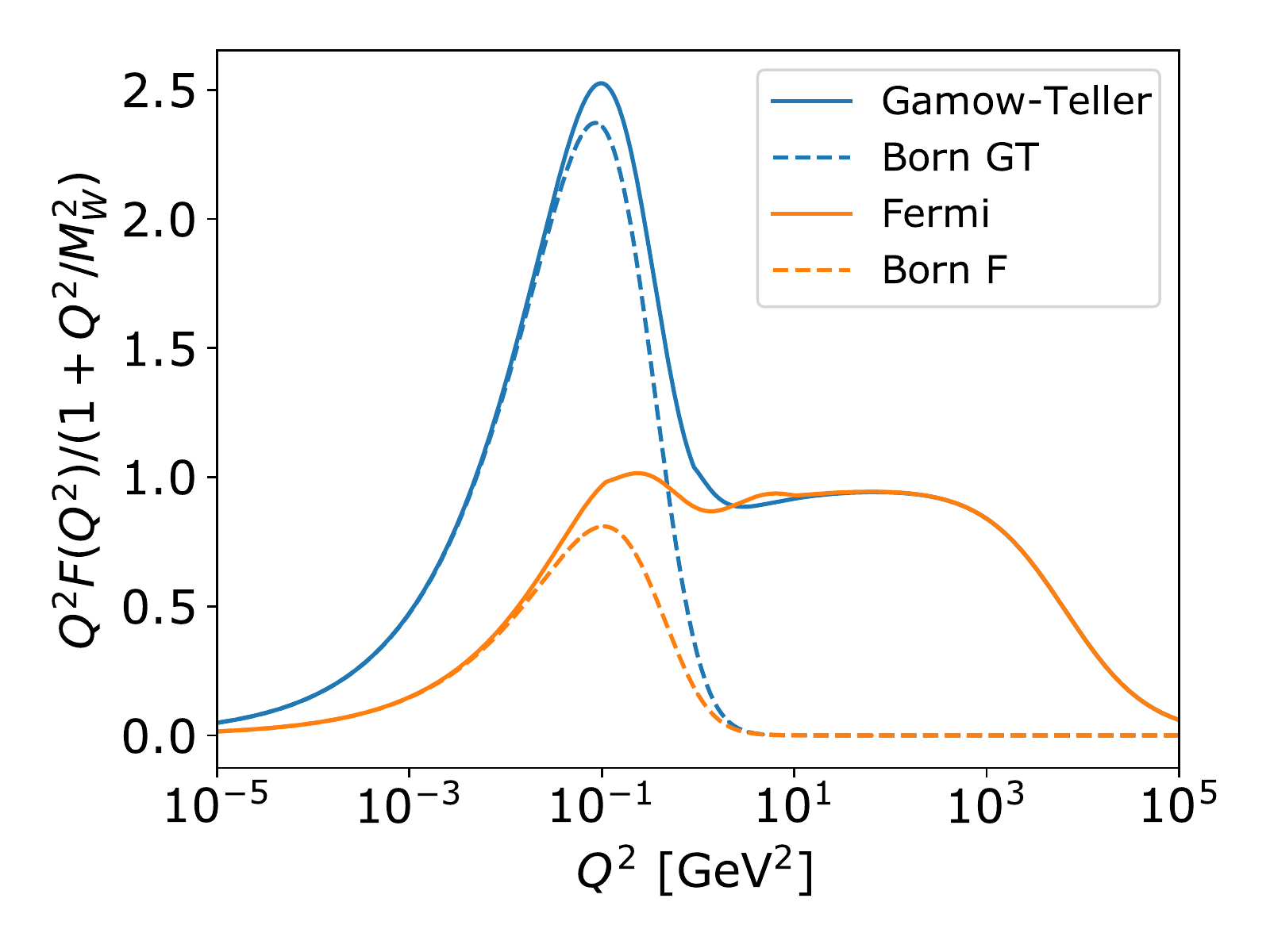}
    \caption{Summary of the results for vector (Fermi) and axial (Gamow-Teller) transitions including target mass corrections, shown as in Ref. \cite{Seng2019b}, using the relationship $F(Q^2) = 12 M_3(1, Q^2)/Q^2$, where $M_3$ is the Nachtmann moment. Dashed lines show the contribution of the Born amplitude.}
    \label{fig:Mellin_Seng}
\end{figure}

So far, an important ingredient is missing from the approach of Eqs. (\ref{eq:PBjSR_DIS})-(\ref{eq:GLS_def}) at low $Q^2$. Since both sum rules are derived at $Q^2 \gg M^2_n$, significant deviations occur when $Q^2$ approaches $M^2$ due to higher-twist (HT) effects: non-perturbative contributions and target mass corrections (TMC) \cite{Schienbein2008, Blumlein1999}. The data shown in Fig. \ref{fig:PBjSr} had TMC removed in order to get clean access to $\alpha_{g_1}(Q^2)$ \cite{Ayala2018a}. In the treatment of Eq. (\ref{eq:T_gamma}), however, such kinematic TMC must necessarily be included. We retroactively apply these corrections by moving to the Nachtmann rather than Mellin moment of Eqs. (\ref{eq:PBjSR_DIS}) and (\ref{eq:GLS_def}). We follow an approach similar to Ref. \cite{Kim1998} and parametrize the sum rule data using power laws and determine the relative shift between Mellin and Nachtmann moments. This produces a correction to Eq. (\ref{eq:sum_rule_contributions})
\begin{equation}
    V_{NP}^\mathrm{TMC} = A_{NP}^\mathrm{TMC} = 0.112(56)\frac{\alpha}{\pi}
\end{equation}
and where we conservatively placed a 50\% relative uncertainty on the correction. Figure \ref{fig:Mellin_Seng} shows a graphical summary of our results, with the area under the curve proportional to $\Delta_R^L$. We then find $\Delta_R^A-\Delta_R^V = 4.07(8) \times 10^{-3}$, which is substantially larger than anticipated. Additionally, the TMC and full integration of the Born contribution fully close the gap between the two most recent $\Delta_R^V$ calculations \cite{Seng2018, Seng2019, Czarnecki2019}.

\begin{figure}[t]
    \centering
    \includegraphics[width=0.48\textwidth]{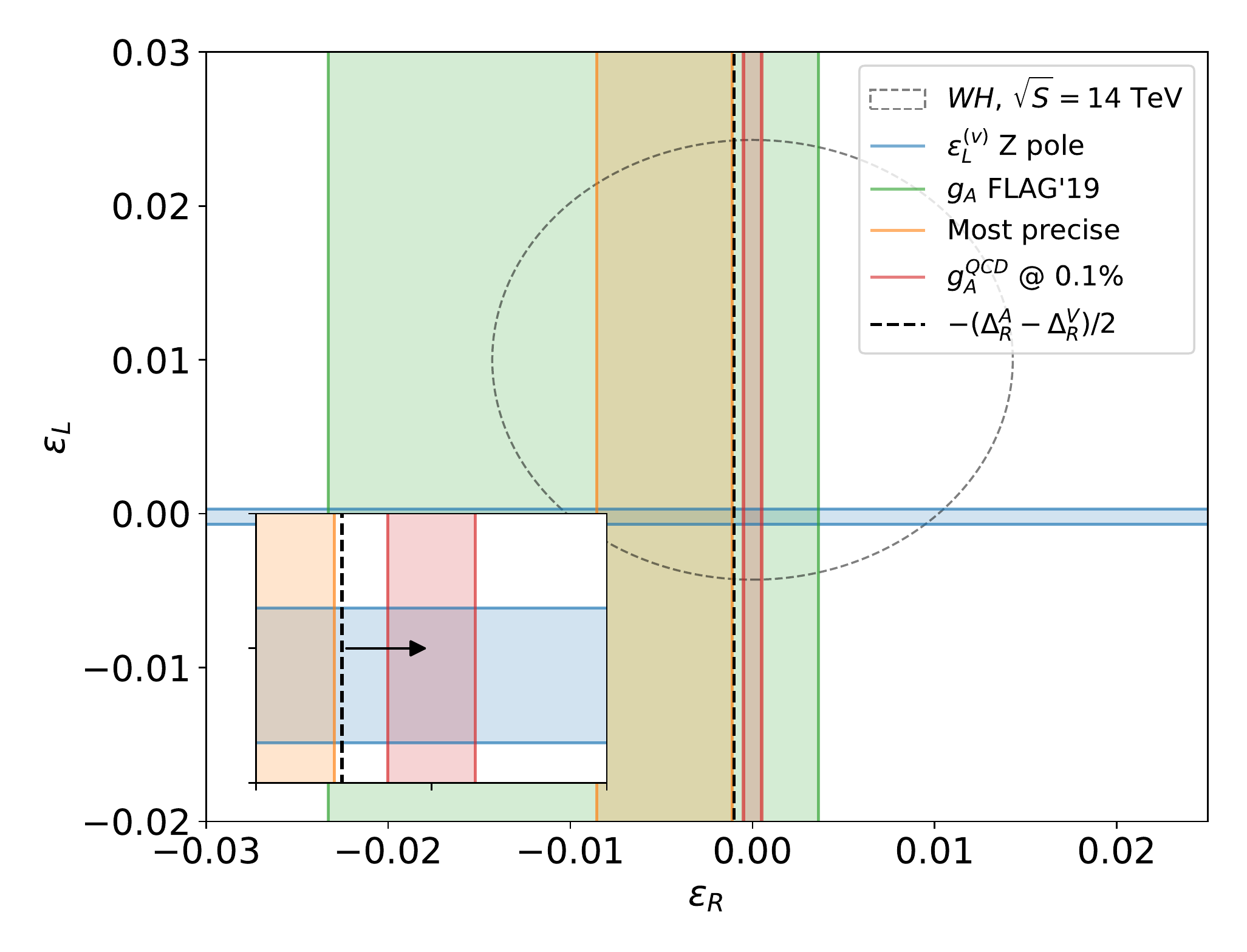}
    \caption{Current limits (68\% C.L.) on left and right-handed couplings interpreted in the SMEFT, showing $Z$-pole (blue) \cite{Falkowski2017, Efrati2015}, LHC (black) \cite{TheATLAScollaboration2016}, LQCD results from FLAG'19 (green) \cite{Aoki2020} and Ref. \cite{Walker-Loud2020} (orange). In red we show anticipated limits when $g_A$ reaches 0.1\% on the lattice. The black vertical line shows $\Delta_R^A-\Delta_R^V$ as a false BSM signal.}
    \label{fig:epsilonR_limits}
\end{figure}

By correcting the experimentally obtained $g_A^\mathrm{eff}$ to find $g_A^0$, we can now make a clean comparison against LQCD results to probe exotic right-handed currents. Using the most precise, published values for $g_A^\mathrm{eff}$ \cite{Markisch2019} and $g_A^\mathrm{LQCD}$ \cite{Walker-Loud2020}, we find $\epsilon_R = -5(4)\times  10^{-3}$, compared to $-10(13)\times 10^{-3}$ when using the FLAG`19 result \cite{Aoki2020}. The results are summarized in Fig. \ref{fig:epsilonR_limits}. When $\delta g_A^\mathrm{QCD}$ reaches 0.1\% on the lattice (i.e. four times as precise as the best published calculation), the effect from $\Delta_R^A-\Delta_R^V$ corresponds to a $2\sigma$ false BSM signal.

\begin{figure}[!ht]
    \centering
    \includegraphics[width=0.48\textwidth]{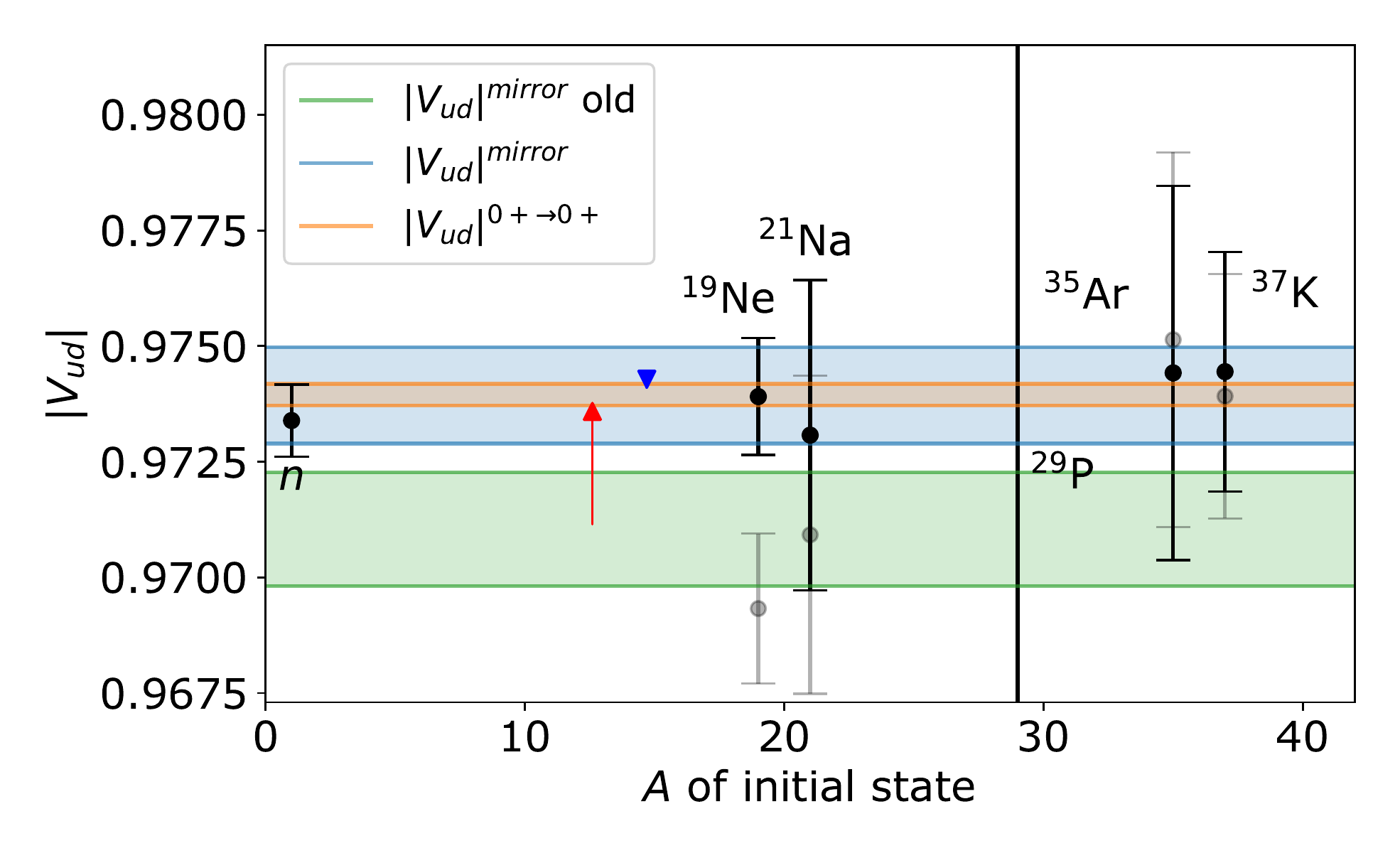}
    \caption{Results with $1\sigma$ uncertainty of $|V_{ud}|$ from mirror decays, superallowed $0^+\to 0^+$ Fermi decays \cite{Hardy2020}, and the neutron. The red arrow shows the central value shift due to the resolved double-counting. Results using uncorrected $(f_A/f_V)$ values for each mirror isotope in gray, with their current value in black.}
    \label{fig:overview_Vud_nuclear}
\end{figure}

Finally, mirror $T=1/2$ decays can be thought of as a nuclear equivalent of the neutron by substituting $3g_A^2 \to \rho^2$ in Eq. (\ref{eq:general_tiple_relation}) and also allow extraction of $|V_{ud}|^\mathrm{mirror}$ \cite{Severijns2008, Naviliat-Cuncic2009}. We resolve a double-counting issue by noticing that the weak magnetism Born contribution of Eq. (\ref{eq:F_A_Born}) is included in the formalism through which one typically calculates the phase space factors \cite{Severijns2008, Towner2015}, $f_{V, A}$, but absent in those for which one extracts $\rho$ from experimental $\beta$-correlation measurements \cite{Holstein1974}. The summary of these changes is shown in Fig. \ref{fig:overview_Vud_nuclear}. The new results solve the long-standing internal discrepancies in the mirror nuclei data set, have a reduced uncertainty and now agree extremely well with both superallowed and neutron data \cite{Hardy2020}.

We conclude by noting that the dominant uncertainty on $\Delta_R^L$ arises from the low momentum behaviour of the QCD sum rules and their target mass corrections, both of which can be tackled using lattice QCD \cite{Gockeler1996}. Even though the precision requirements are more stringent at low $Q^2$ due to due to multiparticle states, recent progress has shown great promise \cite{Chambers2017}. In parallel with improved lattice $g_A$ calculations, our result enables the most stringent constraints on exotic right-handed currents from a clean channel.

\begin{acknowledgements}
I would like to thank Chien-Yeah Seng, Vincenzo Cirigliano, Mikhail Gorchtein, Barry R. Holstein, Nathal Severijns, Albert Young, Andre Walker-Loud and the organizers of ECT*: \textit{Precise beta decay calculations for searches for new physics} and ACFI Amherst: \textit{Current and Future Status of the First-Row CKM Unitarity} workshops for productive discussions related to this manuscript. I acknowledge support by the U.S. National Science Foundation (PHY-1914133), U.S. Department of Energy (DE-FG02-ER41042), the Belgian Federal Science Policy Office (IUAP EP/12-c) and the Fund for Scientific Research Flanders (FWO).
\end{acknowledgements}

\bibliography{library}

\begin{thebibliography}{54}%
\makeatletter
\providecommand \@ifxundefined [1]{%
 \@ifx{#1\undefined}
}%
\providecommand \@ifnum [1]{%
 \ifnum #1\expandafter \@firstoftwo
 \else \expandafter \@secondoftwo
 \fi
}%
\providecommand \@ifx [1]{%
 \ifx #1\expandafter \@firstoftwo
 \else \expandafter \@secondoftwo
 \fi
}%
\providecommand \natexlab [1]{#1}%
\providecommand \enquote  [1]{``#1''}%
\providecommand \bibnamefont  [1]{#1}%
\providecommand \bibfnamefont [1]{#1}%
\providecommand \citenamefont [1]{#1}%
\providecommand \href@noop [0]{\@secondoftwo}%
\providecommand \href [0]{\begingroup \@sanitize@url \@href}%
\providecommand \@href[1]{\@@startlink{#1}\@@href}%
\providecommand \@@href[1]{\endgroup#1\@@endlink}%
\providecommand \@sanitize@url [0]{\catcode `\\12\catcode `\$12\catcode
  `\&12\catcode `\#12\catcode `\^12\catcode `\_12\catcode `\%12\relax}%
\providecommand \@@startlink[1]{}%
\providecommand \@@endlink[0]{}%
\providecommand \url  [0]{\begingroup\@sanitize@url \@url }%
\providecommand \@url [1]{\endgroup\@href {#1}{\urlprefix }}%
\providecommand \urlprefix  [0]{URL }%
\providecommand \Eprint [0]{\href }%
\providecommand \doibase [0]{http://dx.doi.org/}%
\providecommand \selectlanguage [0]{\@gobble}%
\providecommand \bibinfo  [0]{\@secondoftwo}%
\providecommand \bibfield  [0]{\@secondoftwo}%
\providecommand \translation [1]{[#1]}%
\providecommand \BibitemOpen [0]{}%
\providecommand \bibitemStop [0]{}%
\providecommand \bibitemNoStop [0]{.\EOS\space}%
\providecommand \EOS [0]{\spacefactor3000\relax}%
\providecommand \BibitemShut  [1]{\csname bibitem#1\endcsname}%
\let\auto@bib@innerbib\@empty
\bibitem [{\citenamefont {Gonz{\'{a}}lez-Alonso}\ \emph
  {et~al.}(2019)\citenamefont {Gonz{\'{a}}lez-Alonso}, \citenamefont
  {Naviliat-Cuncic},\ and\ \citenamefont {Severijns}}]{Gonzalez-Alonso2018}%
  \BibitemOpen
  \bibfield  {author} {\bibinfo {author} {\bibfnamefont {M.}~\bibnamefont
  {Gonz{\'{a}}lez-Alonso}}, \bibinfo {author} {\bibfnamefont {O.}~\bibnamefont
  {Naviliat-Cuncic}}, \ and\ \bibinfo {author} {\bibfnamefont {N.}~\bibnamefont
  {Severijns}},\ }\href {\doibase 10.1016/j.ppnp.2018.08.002} {\bibfield
  {journal} {\bibinfo  {journal} {Progress in Particle and Nuclear Physics}\
  }\textbf {\bibinfo {volume} {104}},\ \bibinfo {pages} {165} (\bibinfo {year}
  {2019})},\ \Eprint {http://arxiv.org/abs/1803.08732} {arXiv:1803.08732}
  \BibitemShut {NoStop}%
\bibitem [{\citenamefont {Wauters}\ \emph {et~al.}(2014)\citenamefont
  {Wauters}, \citenamefont {Garc{\'{i}}a},\ and\ \citenamefont
  {Hong}}]{Wauters2014}%
  \BibitemOpen
  \bibfield  {author} {\bibinfo {author} {\bibfnamefont {F.}~\bibnamefont
  {Wauters}}, \bibinfo {author} {\bibfnamefont {A.}~\bibnamefont
  {Garc{\'{i}}a}}, \ and\ \bibinfo {author} {\bibfnamefont {R.}~\bibnamefont
  {Hong}},\ }\href {\doibase 10.1103/PhysRevC.89.025501} {\bibfield  {journal}
  {\bibinfo  {journal} {Physical Review C}\ }\textbf {\bibinfo {volume} {89}},\
  \bibinfo {pages} {025501} (\bibinfo {year} {2014})},\ \Eprint
  {http://arxiv.org/abs/1306.2608} {arXiv:1306.2608} \BibitemShut {NoStop}%
\bibitem [{\citenamefont {Vos}\ \emph {et~al.}(2015)\citenamefont {Vos},
  \citenamefont {Wilschut},\ and\ \citenamefont {Timmermans}}]{Vos2015}%
  \BibitemOpen
  \bibfield  {author} {\bibinfo {author} {\bibfnamefont {K.~K.}\ \bibnamefont
  {Vos}}, \bibinfo {author} {\bibfnamefont {H.~W.}\ \bibnamefont {Wilschut}}, \
  and\ \bibinfo {author} {\bibfnamefont {R.~G.~E.}\ \bibnamefont
  {Timmermans}},\ }\href {\doibase 10.1103/RevModPhys.87.1483} {\bibfield
  {journal} {\bibinfo  {journal} {Reviews of Modern Physics}\ }\textbf
  {\bibinfo {volume} {87}},\ \bibinfo {pages} {1483} (\bibinfo {year}
  {2015})},\ \Eprint {http://arxiv.org/abs/1509.04007} {arXiv:1509.04007}
  \BibitemShut {NoStop}%
\bibitem [{\citenamefont {Cirigliano}\ \emph {et~al.}(2013)\citenamefont
  {Cirigliano}, \citenamefont {Gonz{\'{a}}lez-Alonso},\ and\ \citenamefont
  {Graesser}}]{Cirigliano2013}%
  \BibitemOpen
  \bibfield  {author} {\bibinfo {author} {\bibfnamefont {V.}~\bibnamefont
  {Cirigliano}}, \bibinfo {author} {\bibfnamefont {M.}~\bibnamefont
  {Gonz{\'{a}}lez-Alonso}}, \ and\ \bibinfo {author} {\bibfnamefont {M.~L.}\
  \bibnamefont {Graesser}},\ }\href {\doibase 10.1007/JHEP02(2013)046}
  {\bibfield  {journal} {\bibinfo  {journal} {Journal of High Energy Physics}\
  }\textbf {\bibinfo {volume} {2013}},\ \bibinfo {pages} {46} (\bibinfo {year}
  {2013})},\ \Eprint {http://arxiv.org/abs/1210.4553} {arXiv:1210.4553}
  \BibitemShut {NoStop}%
\bibitem [{\citenamefont {Cirigliano}\ and\ \citenamefont
  {Ramsey-Musolf}(2013)}]{Cirigliano2013b}%
  \BibitemOpen
  \bibfield  {author} {\bibinfo {author} {\bibfnamefont {V.}~\bibnamefont
  {Cirigliano}}\ and\ \bibinfo {author} {\bibfnamefont {M.~J.}\ \bibnamefont
  {Ramsey-Musolf}},\ }\href {\doibase 10.1016/j.ppnp.2013.03.002} {\bibfield
  {journal} {\bibinfo  {journal} {Progress in Particle and Nuclear Physics}\
  }\textbf {\bibinfo {volume} {71}},\ \bibinfo {pages} {2} (\bibinfo {year}
  {2013})},\ \Eprint {http://arxiv.org/abs/arXiv:1304.0017v1}
  {arXiv:arXiv:1304.0017v1} \BibitemShut {NoStop}%
\bibitem [{\citenamefont {Pattie}\ \emph {et~al.}(2013)\citenamefont {Pattie},
  \citenamefont {Hickerson},\ and\ \citenamefont {Young}}]{Pattie2013}%
  \BibitemOpen
  \bibfield  {author} {\bibinfo {author} {\bibfnamefont {R.~W.}\ \bibnamefont
  {Pattie}}, \bibinfo {author} {\bibfnamefont {K.~P.}\ \bibnamefont
  {Hickerson}}, \ and\ \bibinfo {author} {\bibfnamefont {A.~R.}\ \bibnamefont
  {Young}},\ }\href {\doibase 10.1103/PhysRevC.88.048501} {\bibfield  {journal}
  {\bibinfo  {journal} {Physical Review C}\ }\textbf {\bibinfo {volume} {88}},\
  \bibinfo {pages} {048501} (\bibinfo {year} {2013})}\BibitemShut {NoStop}%
\bibitem [{\citenamefont {Cirgiliano}\ \emph {et~al.}(2019)\citenamefont
  {Cirgiliano}, \citenamefont {Garcia}, \citenamefont {Gazit}, \citenamefont
  {Naviliat-Cuncic}, \citenamefont {Savard},\ and\ \citenamefont
  {Young}}]{Cirgiliano2019}%
  \BibitemOpen
  \bibfield  {author} {\bibinfo {author} {\bibfnamefont {V.}~\bibnamefont
  {Cirgiliano}}, \bibinfo {author} {\bibfnamefont {A.}~\bibnamefont {Garcia}},
  \bibinfo {author} {\bibfnamefont {D.}~\bibnamefont {Gazit}}, \bibinfo
  {author} {\bibfnamefont {O.}~\bibnamefont {Naviliat-Cuncic}}, \bibinfo
  {author} {\bibfnamefont {G.}~\bibnamefont {Savard}}, \ and\ \bibinfo {author}
  {\bibfnamefont {A.}~\bibnamefont {Young}},\ }\href
  {http://arxiv.org/abs/1907.02164} {\bibfield  {journal} {\bibinfo  {journal}
  {arXiv}\ } (\bibinfo {year} {2019})},\ \Eprint
  {http://arxiv.org/abs/1907.02164} {arXiv:1907.02164} \BibitemShut {NoStop}%
\bibitem [{\citenamefont {Alioli}\ \emph {et~al.}(2017)\citenamefont {Alioli},
  \citenamefont {Cirigliano}, \citenamefont {Dekens}, \citenamefont
  {de~Vries},\ and\ \citenamefont {Mereghetti}}]{Alioli2017}%
  \BibitemOpen
  \bibfield  {author} {\bibinfo {author} {\bibfnamefont {S.}~\bibnamefont
  {Alioli}}, \bibinfo {author} {\bibfnamefont {V.}~\bibnamefont {Cirigliano}},
  \bibinfo {author} {\bibfnamefont {W.}~\bibnamefont {Dekens}}, \bibinfo
  {author} {\bibfnamefont {J.}~\bibnamefont {de~Vries}}, \ and\ \bibinfo
  {author} {\bibfnamefont {E.}~\bibnamefont {Mereghetti}},\ }\href {\doibase
  10.1007/JHEP05(2017)086} {\bibfield  {journal} {\bibinfo  {journal} {Journal
  of High Energy Physics}\ }\textbf {\bibinfo {volume} {2017}},\ \bibinfo
  {pages} {86} (\bibinfo {year} {2017})}\BibitemShut {NoStop}%
\bibitem [{\citenamefont {Sirlin}\ and\ \citenamefont
  {Ferroglia}(2013)}]{Sirlin2013}%
  \BibitemOpen
  \bibfield  {author} {\bibinfo {author} {\bibfnamefont {A.}~\bibnamefont
  {Sirlin}}\ and\ \bibinfo {author} {\bibfnamefont {A.}~\bibnamefont
  {Ferroglia}},\ }\href {\doibase 10.1103/RevModPhys.85.263} {\bibfield
  {journal} {\bibinfo  {journal} {Reviews of Modern Physics}\ }\textbf
  {\bibinfo {volume} {85}},\ \bibinfo {pages} {263} (\bibinfo {year} {2013})},\
  \Eprint {http://arxiv.org/abs/1210.5296} {arXiv:1210.5296} \BibitemShut
  {NoStop}%
\bibitem [{\citenamefont {Czarnecki}\ \emph {et~al.}(2018)\citenamefont
  {Czarnecki}, \citenamefont {Marciano},\ and\ \citenamefont
  {Sirlin}}]{Czarnecki2018}%
  \BibitemOpen
  \bibfield  {author} {\bibinfo {author} {\bibfnamefont {A.}~\bibnamefont
  {Czarnecki}}, \bibinfo {author} {\bibfnamefont {W.~J.}\ \bibnamefont
  {Marciano}}, \ and\ \bibinfo {author} {\bibfnamefont {A.}~\bibnamefont
  {Sirlin}},\ }\href {\doibase 10.1103/PhysRevLett.120.202002} {\bibfield
  {journal} {\bibinfo  {journal} {Physical Review Letters}\ }\textbf {\bibinfo
  {volume} {120}},\ \bibinfo {pages} {202002} (\bibinfo {year} {2018})},\
  \Eprint {http://arxiv.org/abs/1802.01804} {arXiv:1802.01804} \BibitemShut
  {NoStop}%
\bibitem [{\citenamefont {Ademollo}\ and\ \citenamefont
  {Gatto}(1964)}]{Ademollo1964}%
  \BibitemOpen
  \bibfield  {author} {\bibinfo {author} {\bibfnamefont {M.}~\bibnamefont
  {Ademollo}}\ and\ \bibinfo {author} {\bibfnamefont {R.}~\bibnamefont
  {Gatto}},\ }\href {\doibase 10.1103/PhysRevLett.13.264} {\bibfield  {journal}
  {\bibinfo  {journal} {Physical Review Letters}\ }\textbf {\bibinfo {volume}
  {13}},\ \bibinfo {pages} {264} (\bibinfo {year} {1964})}\BibitemShut
  {NoStop}%
\bibitem [{\citenamefont {Sirlin}(1978)}]{Sirlin1978}%
  \BibitemOpen
  \bibfield  {author} {\bibinfo {author} {\bibfnamefont {A.}~\bibnamefont
  {Sirlin}},\ }\href {\doibase 10.1088/0305-4470/36/15/304} {\bibfield
  {journal} {\bibinfo  {journal} {Reviews of Modern Physics}\ }\textbf
  {\bibinfo {volume} {50}},\ \bibinfo {pages} {573} (\bibinfo {year}
  {1978})}\BibitemShut {NoStop}%
\bibitem [{\citenamefont {Seng}\ \emph {et~al.}(2018)\citenamefont {Seng},
  \citenamefont {Gorchtein}, \citenamefont {Patel},\ and\ \citenamefont
  {Ramsey-Musolf}}]{Seng2018}%
  \BibitemOpen
  \bibfield  {author} {\bibinfo {author} {\bibfnamefont {C.-Y.}\ \bibnamefont
  {Seng}}, \bibinfo {author} {\bibfnamefont {M.}~\bibnamefont {Gorchtein}},
  \bibinfo {author} {\bibfnamefont {H.~H.}\ \bibnamefont {Patel}}, \ and\
  \bibinfo {author} {\bibfnamefont {M.~J.}\ \bibnamefont {Ramsey-Musolf}},\
  }\href {\doibase 10.1103/PhysRevLett.121.241804} {\bibfield  {journal}
  {\bibinfo  {journal} {Physical Review Letters}\ }\textbf {\bibinfo {volume}
  {121}},\ \bibinfo {pages} {241804} (\bibinfo {year} {2018})},\ \Eprint
  {http://arxiv.org/abs/1807.10197} {arXiv:1807.10197} \BibitemShut {NoStop}%
\bibitem [{\citenamefont {Seng}\ \emph {et~al.}(2020)\citenamefont {Seng},
  \citenamefont {Galviz},\ and\ \citenamefont {Mei{\ss}ner}}]{Seng2019}%
  \BibitemOpen
  \bibfield  {author} {\bibinfo {author} {\bibfnamefont {C.-Y.}\ \bibnamefont
  {Seng}}, \bibinfo {author} {\bibfnamefont {D.}~\bibnamefont {Galviz}}, \ and\
  \bibinfo {author} {\bibfnamefont {U.-G.}\ \bibnamefont {Mei{\ss}ner}},\
  }\href {\doibase 10.1007/JHEP02(2020)069} {\bibfield  {journal} {\bibinfo
  {journal} {Journal of High Energy Physics}\ }\textbf {\bibinfo {volume}
  {2020}},\ \bibinfo {pages} {69} (\bibinfo {year} {2020})},\ \Eprint
  {http://arxiv.org/abs/1910.13208} {arXiv:1910.13208} \BibitemShut {NoStop}%
\bibitem [{\citenamefont {Czarnecki}\ \emph {et~al.}(2019)\citenamefont
  {Czarnecki}, \citenamefont {Marciano},\ and\ \citenamefont
  {Sirlin}}]{Czarnecki2019}%
  \BibitemOpen
  \bibfield  {author} {\bibinfo {author} {\bibfnamefont {A.}~\bibnamefont
  {Czarnecki}}, \bibinfo {author} {\bibfnamefont {W.~J.}\ \bibnamefont
  {Marciano}}, \ and\ \bibinfo {author} {\bibfnamefont {A.}~\bibnamefont
  {Sirlin}},\ }\href {\doibase 10.1103/PhysRevD.100.073008} {\bibfield
  {journal} {\bibinfo  {journal} {Physical Review D}\ }\textbf {\bibinfo
  {volume} {100}},\ \bibinfo {pages} {73008} (\bibinfo {year} {2019})},\
  \Eprint {http://arxiv.org/abs/1907.06737} {arXiv:1907.06737} \BibitemShut
  {NoStop}%
\bibitem [{\citenamefont {Bhattacharya}\ \emph {et~al.}(2012)\citenamefont
  {Bhattacharya}, \citenamefont {Cirigliano}, \citenamefont {Cohen},
  \citenamefont {Filipuzzi}, \citenamefont {Gonz{\'{a}}lez-Alonso},
  \citenamefont {Graesser}, \citenamefont {Gupta},\ and\ \citenamefont
  {Lin}}]{Bhattacharya2012}%
  \BibitemOpen
  \bibfield  {author} {\bibinfo {author} {\bibfnamefont {T.}~\bibnamefont
  {Bhattacharya}}, \bibinfo {author} {\bibfnamefont {V.}~\bibnamefont
  {Cirigliano}}, \bibinfo {author} {\bibfnamefont {S.~D.}\ \bibnamefont
  {Cohen}}, \bibinfo {author} {\bibfnamefont {A.}~\bibnamefont {Filipuzzi}},
  \bibinfo {author} {\bibfnamefont {M.}~\bibnamefont {Gonz{\'{a}}lez-Alonso}},
  \bibinfo {author} {\bibfnamefont {M.~L.}\ \bibnamefont {Graesser}}, \bibinfo
  {author} {\bibfnamefont {R.}~\bibnamefont {Gupta}}, \ and\ \bibinfo {author}
  {\bibfnamefont {H.-W.}\ \bibnamefont {Lin}},\ }\href {\doibase
  10.1103/PhysRevD.85.054512} {\bibfield  {journal} {\bibinfo  {journal}
  {Physical Review D}\ }\textbf {\bibinfo {volume} {85}},\ \bibinfo {pages}
  {054512} (\bibinfo {year} {2012})},\ \Eprint {http://arxiv.org/abs/1110.6448}
  {arXiv:1110.6448} \BibitemShut {NoStop}%
\bibitem [{\citenamefont {Gupta}\ \emph {et~al.}(2018)\citenamefont {Gupta},
  \citenamefont {Jang}, \citenamefont {Yoon}, \citenamefont {Lin},
  \citenamefont {Cirigliano},\ and\ \citenamefont {Bhattacharya}}]{Gupta2018}%
  \BibitemOpen
  \bibfield  {author} {\bibinfo {author} {\bibfnamefont {R.}~\bibnamefont
  {Gupta}}, \bibinfo {author} {\bibfnamefont {Y.-C.}\ \bibnamefont {Jang}},
  \bibinfo {author} {\bibfnamefont {B.}~\bibnamefont {Yoon}}, \bibinfo {author}
  {\bibfnamefont {H.-W.}\ \bibnamefont {Lin}}, \bibinfo {author} {\bibfnamefont
  {V.}~\bibnamefont {Cirigliano}}, \ and\ \bibinfo {author} {\bibfnamefont
  {T.}~\bibnamefont {Bhattacharya}},\ }\href {\doibase
  10.1103/PhysRevD.98.034503} {\bibfield  {journal} {\bibinfo  {journal}
  {Physical Review D}\ }\textbf {\bibinfo {volume} {98}},\ \bibinfo {pages}
  {034503} (\bibinfo {year} {2018})},\ \Eprint
  {http://arxiv.org/abs/1806.09006} {arXiv:1806.09006} \BibitemShut {NoStop}%
\bibitem [{\citenamefont {Chang}\ \emph {et~al.}(2018)\citenamefont {Chang},
  \citenamefont {Nicholson}, \citenamefont {Rinaldi}, \citenamefont
  {Berkowitz}, \citenamefont {Garron}, \citenamefont {Brantley}, \citenamefont
  {Monge-Camacho}, \citenamefont {Monahan}, \citenamefont {Bouchard},
  \citenamefont {Clark}, \citenamefont {Jo{\'{o}}}, \citenamefont {Kurth},
  \citenamefont {Orginos}, \citenamefont {Vranas},\ and\ \citenamefont
  {Walker-Loud}}]{Chang2018}%
  \BibitemOpen
  \bibfield  {author} {\bibinfo {author} {\bibfnamefont {C.~C.}\ \bibnamefont
  {Chang}}, \bibinfo {author} {\bibfnamefont {A.~N.}\ \bibnamefont
  {Nicholson}}, \bibinfo {author} {\bibfnamefont {E.}~\bibnamefont {Rinaldi}},
  \bibinfo {author} {\bibfnamefont {E.}~\bibnamefont {Berkowitz}}, \bibinfo
  {author} {\bibfnamefont {N.}~\bibnamefont {Garron}}, \bibinfo {author}
  {\bibfnamefont {D.~A.}\ \bibnamefont {Brantley}}, \bibinfo {author}
  {\bibfnamefont {H.}~\bibnamefont {Monge-Camacho}}, \bibinfo {author}
  {\bibfnamefont {C.~J.}\ \bibnamefont {Monahan}}, \bibinfo {author}
  {\bibfnamefont {C.}~\bibnamefont {Bouchard}}, \bibinfo {author}
  {\bibfnamefont {M.~A.}\ \bibnamefont {Clark}}, \bibinfo {author}
  {\bibfnamefont {B.}~\bibnamefont {Jo{\'{o}}}}, \bibinfo {author}
  {\bibfnamefont {T.}~\bibnamefont {Kurth}}, \bibinfo {author} {\bibfnamefont
  {K.}~\bibnamefont {Orginos}}, \bibinfo {author} {\bibfnamefont
  {P.}~\bibnamefont {Vranas}}, \ and\ \bibinfo {author} {\bibfnamefont
  {A.}~\bibnamefont {Walker-Loud}},\ }\href {\doibase
  10.1038/s41586-018-0161-8} {\bibfield  {journal} {\bibinfo  {journal}
  {Nature}\ }\textbf {\bibinfo {volume} {558}},\ \bibinfo {pages} {91}
  (\bibinfo {year} {2018})},\ \Eprint {http://arxiv.org/abs/1805.12130}
  {arXiv:1805.12130} \BibitemShut {NoStop}%
\bibitem [{\citenamefont {Aoki}\ \emph {et~al.}(2020)\citenamefont {Aoki},
  \citenamefont {Aoki}, \citenamefont {Be{\v{c}}irevi{\'{c}}}, \citenamefont
  {Blum}, \citenamefont {Colangelo}, \citenamefont {Collins}, \citenamefont
  {{Della Morte}}, \citenamefont {Dimopoulos}, \citenamefont {D{\"{u}}rr},
  \citenamefont {Fukaya}, \citenamefont {Golterman}, \citenamefont {Gottlieb},
  \citenamefont {Gupta}, \citenamefont {Hashimoto}, \citenamefont {Heller},
  \citenamefont {Herdoiza}, \citenamefont {Horsley}, \citenamefont
  {J{\"{u}}ttner}, \citenamefont {Kaneko}, \citenamefont {Lin}, \citenamefont
  {Lunghi}, \citenamefont {Mawhinney}, \citenamefont {Nicholson}, \citenamefont
  {Onogi}, \citenamefont {Pena}, \citenamefont {Portelli}, \citenamefont
  {Ramos}, \citenamefont {Sharpe}, \citenamefont {Simone}, \citenamefont
  {Simula}, \citenamefont {Sommer}, \citenamefont {{Van de Water}},
  \citenamefont {Vladikas}, \citenamefont {Wenger},\ and\ \citenamefont
  {Wittig}}]{Aoki2020}%
  \BibitemOpen
  \bibfield  {author} {\bibinfo {author} {\bibfnamefont {S.}~\bibnamefont
  {Aoki}}, \bibinfo {author} {\bibfnamefont {Y.}~\bibnamefont {Aoki}}, \bibinfo
  {author} {\bibfnamefont {D.}~\bibnamefont {Be{\v{c}}irevi{\'{c}}}}, \bibinfo
  {author} {\bibfnamefont {T.}~\bibnamefont {Blum}}, \bibinfo {author}
  {\bibfnamefont {G.}~\bibnamefont {Colangelo}}, \bibinfo {author}
  {\bibfnamefont {S.}~\bibnamefont {Collins}}, \bibinfo {author} {\bibfnamefont
  {M.}~\bibnamefont {{Della Morte}}}, \bibinfo {author} {\bibfnamefont
  {P.}~\bibnamefont {Dimopoulos}}, \bibinfo {author} {\bibfnamefont
  {S.}~\bibnamefont {D{\"{u}}rr}}, \bibinfo {author} {\bibfnamefont
  {H.}~\bibnamefont {Fukaya}}, \bibinfo {author} {\bibfnamefont
  {M.}~\bibnamefont {Golterman}}, \bibinfo {author} {\bibfnamefont
  {S.}~\bibnamefont {Gottlieb}}, \bibinfo {author} {\bibfnamefont
  {R.}~\bibnamefont {Gupta}}, \bibinfo {author} {\bibfnamefont
  {S.}~\bibnamefont {Hashimoto}}, \bibinfo {author} {\bibfnamefont {U.~M.}\
  \bibnamefont {Heller}}, \bibinfo {author} {\bibfnamefont {G.}~\bibnamefont
  {Herdoiza}}, \bibinfo {author} {\bibfnamefont {R.}~\bibnamefont {Horsley}},
  \bibinfo {author} {\bibfnamefont {A.}~\bibnamefont {J{\"{u}}ttner}}, \bibinfo
  {author} {\bibfnamefont {T.}~\bibnamefont {Kaneko}}, \bibinfo {author}
  {\bibfnamefont {C.-J.~D.}\ \bibnamefont {Lin}}, \bibinfo {author}
  {\bibfnamefont {E.}~\bibnamefont {Lunghi}}, \bibinfo {author} {\bibfnamefont
  {R.}~\bibnamefont {Mawhinney}}, \bibinfo {author} {\bibfnamefont
  {A.}~\bibnamefont {Nicholson}}, \bibinfo {author} {\bibfnamefont
  {T.}~\bibnamefont {Onogi}}, \bibinfo {author} {\bibfnamefont
  {C.}~\bibnamefont {Pena}}, \bibinfo {author} {\bibfnamefont {A.}~\bibnamefont
  {Portelli}}, \bibinfo {author} {\bibfnamefont {A.}~\bibnamefont {Ramos}},
  \bibinfo {author} {\bibfnamefont {S.~R.}\ \bibnamefont {Sharpe}}, \bibinfo
  {author} {\bibfnamefont {J.~N.}\ \bibnamefont {Simone}}, \bibinfo {author}
  {\bibfnamefont {S.}~\bibnamefont {Simula}}, \bibinfo {author} {\bibfnamefont
  {R.}~\bibnamefont {Sommer}}, \bibinfo {author} {\bibfnamefont
  {R.}~\bibnamefont {{Van de Water}}}, \bibinfo {author} {\bibfnamefont
  {A.}~\bibnamefont {Vladikas}}, \bibinfo {author} {\bibfnamefont
  {U.}~\bibnamefont {Wenger}}, \ and\ \bibinfo {author} {\bibfnamefont
  {H.}~\bibnamefont {Wittig}},\ }\href {\doibase
  10.1140/epjc/s10052-019-7354-7} {\bibfield  {journal} {\bibinfo  {journal}
  {The European Physical Journal C}\ }\textbf {\bibinfo {volume} {80}},\
  \bibinfo {pages} {113} (\bibinfo {year} {2020})},\ \Eprint
  {http://arxiv.org/abs/1902.08191} {arXiv:1902.08191} \BibitemShut {NoStop}%
\bibitem [{\citenamefont {Group}(2020)}]{Zyla2020}%
  \BibitemOpen
  \bibfield  {author} {\bibinfo {author} {\bibfnamefont {P.~D.}\ \bibnamefont
  {Group}},\ }\href {\doibase 10.1093/ptep/ptaa104} {\bibfield  {journal}
  {\bibinfo  {journal} {Progress of Theoretical and Experimental Physics}\
  }\textbf {\bibinfo {volume} {2020}} (\bibinfo {year} {2020}),\
  10.1093/ptep/ptaa104}\BibitemShut {NoStop}%
\bibitem [{\citenamefont {Hardy}\ and\ \citenamefont
  {Towner}(2020)}]{Hardy2020}%
  \BibitemOpen
  \bibfield  {author} {\bibinfo {author} {\bibfnamefont {J.~C.}\ \bibnamefont
  {Hardy}}\ and\ \bibinfo {author} {\bibfnamefont {I.~S.}\ \bibnamefont
  {Towner}},\ }\href {\doibase 10.1103/PhysRevC.102.045501} {\bibfield
  {journal} {\bibinfo  {journal} {Physical Review C}\ }\textbf {\bibinfo
  {volume} {102}},\ \bibinfo {pages} {045501} (\bibinfo {year}
  {2020})}\BibitemShut {NoStop}%
\bibitem [{\citenamefont {Hayen}(2020)}]{Hayen2020c}%
  \BibitemOpen
  \bibfield  {author} {\bibinfo {author} {\bibfnamefont {L.}~\bibnamefont
  {Hayen}},\ }\href@noop {} {\  (\bibinfo {year} {2020})},\ \Eprint
  {http://arxiv.org/abs/2010.07262v1} {arXiv:2010.07262v1} \BibitemShut
  {NoStop}%
\bibitem [{\citenamefont {Czarnecki}\ \emph {et~al.}(2004)\citenamefont
  {Czarnecki}, \citenamefont {Marciano},\ and\ \citenamefont
  {Sirlin}}]{Czarnecki2004}%
  \BibitemOpen
  \bibfield  {author} {\bibinfo {author} {\bibfnamefont {A.}~\bibnamefont
  {Czarnecki}}, \bibinfo {author} {\bibfnamefont {W.~J.}\ \bibnamefont
  {Marciano}}, \ and\ \bibinfo {author} {\bibfnamefont {A.}~\bibnamefont
  {Sirlin}},\ }\href {\doibase 10.1103/PhysRevD.70.093006} {\bibfield
  {journal} {\bibinfo  {journal} {Physical Review D - Particles, Fields,
  Gravitation and Cosmology}\ }\textbf {\bibinfo {volume} {70}},\ \bibinfo
  {pages} {093006} (\bibinfo {year} {2004})},\ \Eprint
  {http://arxiv.org/abs/0406324} {arXiv:0406324 [hep-ph]} \BibitemShut
  {NoStop}%
\bibitem [{\citenamefont {Sirlin}(1982)}]{Sirlin1982}%
  \BibitemOpen
  \bibfield  {author} {\bibinfo {author} {\bibfnamefont {A.}~\bibnamefont
  {Sirlin}},\ }\href@noop {} {\bibfield  {journal} {\bibinfo  {journal}
  {Nuclear Physics B}\ }\textbf {\bibinfo {volume} {196}},\ \bibinfo {pages}
  {83} (\bibinfo {year} {1982})}\BibitemShut {NoStop}%
\bibitem [{\citenamefont {Adler}\ and\ \citenamefont
  {Dashen}(1968)}]{Adler1968}%
  \BibitemOpen
  \bibfield  {author} {\bibinfo {author} {\bibfnamefont {S.~L.}\ \bibnamefont
  {Adler}}\ and\ \bibinfo {author} {\bibfnamefont {R.~F.}\ \bibnamefont
  {Dashen}},\ }\href@noop {} {\emph {\bibinfo {title} {{Current Algebras and
  Applications to Particle Physics}}}}\ (\bibinfo  {publisher} {W. A. Benjamin,
  Inc.},\ \bibinfo {address} {New York, NY},\ \bibinfo {year}
  {1968})\BibitemShut {NoStop}%
\bibitem [{\citenamefont {Treiman}\ \emph {et~al.}(1972)\citenamefont
  {Treiman}, \citenamefont {Jackiw},\ and\ \citenamefont
  {Gross}}]{Treiman1972}%
  \BibitemOpen
  \bibfield  {author} {\bibinfo {author} {\bibfnamefont {S.~B.}\ \bibnamefont
  {Treiman}}, \bibinfo {author} {\bibfnamefont {R.}~\bibnamefont {Jackiw}}, \
  and\ \bibinfo {author} {\bibfnamefont {D.~J.}\ \bibnamefont {Gross}},\
  }\href@noop {} {\emph {\bibinfo {title} {{Lectures on Current Algebra and Its
  Applications}}}}\ (\bibinfo  {publisher} {Princeton University Press},\
  \bibinfo {year} {1972})\BibitemShut {NoStop}%
\bibitem [{\citenamefont {Bjorken}(1966)}]{Bjorken1966}%
  \BibitemOpen
  \bibfield  {author} {\bibinfo {author} {\bibfnamefont {J.~D.}\ \bibnamefont
  {Bjorken}},\ }\href {\doibase 10.1103/PhysRev.148.1467} {\bibfield  {journal}
  {\bibinfo  {journal} {Physical Review}\ }\textbf {\bibinfo {volume} {148}},\
  \bibinfo {pages} {1467} (\bibinfo {year} {1966})}\BibitemShut {NoStop}%
\bibitem [{\citenamefont {Johnson}\ \emph {et~al.}(1961)\citenamefont
  {Johnson}, \citenamefont {O'Connell},\ and\ \citenamefont
  {Mullin}}]{Johnson1961}%
  \BibitemOpen
  \bibfield  {author} {\bibinfo {author} {\bibfnamefont {W.~R.}\ \bibnamefont
  {Johnson}}, \bibinfo {author} {\bibfnamefont {R.~F.}\ \bibnamefont
  {O'Connell}}, \ and\ \bibinfo {author} {\bibfnamefont {C.~J.}\ \bibnamefont
  {Mullin}},\ }\href@noop {} {\bibfield  {journal} {\bibinfo  {journal}
  {Physical Review}\ }\textbf {\bibinfo {volume} {124}},\ \bibinfo {pages}
  {904} (\bibinfo {year} {1961})}\BibitemShut {NoStop}%
\bibitem [{\citenamefont {Sirlin}(1968)}]{Sirlin1968}%
  \BibitemOpen
  \bibfield  {author} {\bibinfo {author} {\bibfnamefont {A.}~\bibnamefont
  {Sirlin}},\ }\href {\doibase 10.1103/PhysRev.176.1871} {\bibfield  {journal}
  {\bibinfo  {journal} {Physical Review}\ }\textbf {\bibinfo {volume} {176}},\
  \bibinfo {pages} {1871} (\bibinfo {year} {1968})}\BibitemShut {NoStop}%
\bibitem [{\citenamefont {Weinberg}(1958)}]{Weinberg1958}%
  \BibitemOpen
  \bibfield  {author} {\bibinfo {author} {\bibfnamefont {S.}~\bibnamefont
  {Weinberg}},\ }\href {\doibase 10.1103/PhysRev.112.1375} {\bibfield
  {journal} {\bibinfo  {journal} {Physical Review}\ }\textbf {\bibinfo {volume}
  {112}},\ \bibinfo {pages} {1375} (\bibinfo {year} {1958})}\BibitemShut
  {NoStop}%
\bibitem [{\citenamefont {Marciano}\ and\ \citenamefont
  {Sirlin}(2006)}]{Marciano2006}%
  \BibitemOpen
  \bibfield  {author} {\bibinfo {author} {\bibfnamefont {W.~J.}\ \bibnamefont
  {Marciano}}\ and\ \bibinfo {author} {\bibfnamefont {A.}~\bibnamefont
  {Sirlin}},\ }\href {\doibase 10.1103/PhysRevLett.96.032002} {\bibfield
  {journal} {\bibinfo  {journal} {Physical Review Letters}\ }\textbf {\bibinfo
  {volume} {96}},\ \bibinfo {pages} {032002} (\bibinfo {year} {2006})},\
  \Eprint {http://arxiv.org/abs/0510099} {arXiv:0510099 [hep-ph]} \BibitemShut
  {NoStop}%
\bibitem [{\citenamefont {Sirlin}(1967)}]{Sirlin1967}%
  \BibitemOpen
  \bibfield  {author} {\bibinfo {author} {\bibfnamefont {A.}~\bibnamefont
  {Sirlin}},\ }\href {\doibase 10.1103/PhysRev.164.1767} {\bibfield  {journal}
  {\bibinfo  {journal} {Physical Review}\ }\textbf {\bibinfo {volume} {164}},\
  \bibinfo {pages} {1767} (\bibinfo {year} {1967})}\BibitemShut {NoStop}%
\bibitem [{\citenamefont {Ye}\ \emph {et~al.}(2018)\citenamefont {Ye},
  \citenamefont {Arrington}, \citenamefont {Hill},\ and\ \citenamefont
  {Lee}}]{Ye2018}%
  \BibitemOpen
  \bibfield  {author} {\bibinfo {author} {\bibfnamefont {Z.}~\bibnamefont
  {Ye}}, \bibinfo {author} {\bibfnamefont {J.}~\bibnamefont {Arrington}},
  \bibinfo {author} {\bibfnamefont {R.~J.}\ \bibnamefont {Hill}}, \ and\
  \bibinfo {author} {\bibfnamefont {G.}~\bibnamefont {Lee}},\ }\href {\doibase
  10.1016/j.physletb.2017.11.023} {\bibfield  {journal} {\bibinfo  {journal}
  {Physics Letters B}\ }\textbf {\bibinfo {volume} {777}},\ \bibinfo {pages}
  {8} (\bibinfo {year} {2018})}\BibitemShut {NoStop}%
\bibitem [{\citenamefont {Bhattacharya}\ \emph {et~al.}(2011)\citenamefont
  {Bhattacharya}, \citenamefont {Hill},\ and\ \citenamefont
  {Paz}}]{Bhattacharya2011}%
  \BibitemOpen
  \bibfield  {author} {\bibinfo {author} {\bibfnamefont {B.}~\bibnamefont
  {Bhattacharya}}, \bibinfo {author} {\bibfnamefont {R.~J.}\ \bibnamefont
  {Hill}}, \ and\ \bibinfo {author} {\bibfnamefont {G.}~\bibnamefont {Paz}},\
  }\href {\doibase 10.1103/PhysRevD.84.073006} {\bibfield  {journal} {\bibinfo
  {journal} {Physical Review D}\ }\textbf {\bibinfo {volume} {84}},\ \bibinfo
  {pages} {073006} (\bibinfo {year} {2011})}\BibitemShut {NoStop}%
\bibitem [{\citenamefont {Ayala}\ \emph {et~al.}(2018)\citenamefont {Ayala},
  \citenamefont {Cveti{\v{c}}}, \citenamefont {Kotikov},\ and\ \citenamefont
  {Shaikhatdenov}}]{Ayala2018a}%
  \BibitemOpen
  \bibfield  {author} {\bibinfo {author} {\bibfnamefont {C.}~\bibnamefont
  {Ayala}}, \bibinfo {author} {\bibfnamefont {G.}~\bibnamefont {Cveti{\v{c}}}},
  \bibinfo {author} {\bibfnamefont {A.~V.}\ \bibnamefont {Kotikov}}, \ and\
  \bibinfo {author} {\bibfnamefont {B.~G.}\ \bibnamefont {Shaikhatdenov}},\
  }\href {\doibase 10.1140/epjc/s10052-018-6490-9} {\bibfield  {journal}
  {\bibinfo  {journal} {The European Physical Journal C}\ }\textbf {\bibinfo
  {volume} {78}},\ \bibinfo {pages} {1002} (\bibinfo {year}
  {2018})}\BibitemShut {NoStop}%
\bibitem [{\citenamefont {Baikov}\ \emph {et~al.}(2010)\citenamefont {Baikov},
  \citenamefont {Chetyrkin},\ and\ \citenamefont {K{\"{u}}hn}}]{Baikov2010}%
  \BibitemOpen
  \bibfield  {author} {\bibinfo {author} {\bibfnamefont {P.~A.}\ \bibnamefont
  {Baikov}}, \bibinfo {author} {\bibfnamefont {K.~G.}\ \bibnamefont
  {Chetyrkin}}, \ and\ \bibinfo {author} {\bibfnamefont {J.~H.}\ \bibnamefont
  {K{\"{u}}hn}},\ }\href {\doibase 10.1103/PhysRevLett.104.132004} {\bibfield
  {journal} {\bibinfo  {journal} {Physical Review Letters}\ }\textbf {\bibinfo
  {volume} {104}},\ \bibinfo {pages} {132004} (\bibinfo {year}
  {2010})}\BibitemShut {NoStop}%
\bibitem [{\citenamefont {Brodsky}\ \emph {et~al.}(2010)\citenamefont
  {Brodsky}, \citenamefont {de~T{\'{e}}ramond},\ and\ \citenamefont
  {Deur}}]{Brodsky2010}%
  \BibitemOpen
  \bibfield  {author} {\bibinfo {author} {\bibfnamefont {S.~J.}\ \bibnamefont
  {Brodsky}}, \bibinfo {author} {\bibfnamefont {G.~F.}\ \bibnamefont
  {de~T{\'{e}}ramond}}, \ and\ \bibinfo {author} {\bibfnamefont
  {A.}~\bibnamefont {Deur}},\ }\href {\doibase 10.1103/PhysRevD.81.096010}
  {\bibfield  {journal} {\bibinfo  {journal} {Physical Review D}\ }\textbf
  {\bibinfo {volume} {81}},\ \bibinfo {pages} {096010} (\bibinfo {year}
  {2010})}\BibitemShut {NoStop}%
\bibitem [{\citenamefont {Herren}\ and\ \citenamefont
  {Steinhauser}(2018)}]{Herren2018}%
  \BibitemOpen
  \bibfield  {author} {\bibinfo {author} {\bibfnamefont {F.}~\bibnamefont
  {Herren}}\ and\ \bibinfo {author} {\bibfnamefont {M.}~\bibnamefont
  {Steinhauser}},\ }\href {\doibase 10.1016/j.cpc.2017.11.014} {\bibfield
  {journal} {\bibinfo  {journal} {Computer Physics Communications}\ }\textbf
  {\bibinfo {volume} {224}},\ \bibinfo {pages} {333} (\bibinfo {year}
  {2018})}\BibitemShut {NoStop}%
\bibitem [{\citenamefont {Larin}\ and\ \citenamefont
  {Vermaseren}(1991)}]{Larin1991}%
  \BibitemOpen
  \bibfield  {author} {\bibinfo {author} {\bibfnamefont {S.}~\bibnamefont
  {Larin}}\ and\ \bibinfo {author} {\bibfnamefont {J.}~\bibnamefont
  {Vermaseren}},\ }\href {\doibase 10.1016/0370-2693(91)90839-I} {\bibfield
  {journal} {\bibinfo  {journal} {Physics Letters B}\ }\textbf {\bibinfo
  {volume} {259}},\ \bibinfo {pages} {345} (\bibinfo {year}
  {1991})}\BibitemShut {NoStop}%
\bibitem [{\citenamefont {Kim}\ \emph {et~al.}(1998)\citenamefont {Kim},
  \citenamefont {Harris}, \citenamefont {Arroyo}, \citenamefont {de~Barbaro},
  \citenamefont {de~Barbaro}, \citenamefont {Bazarko}, \citenamefont
  {Bernstein}, \citenamefont {Bodek}, \citenamefont {Bolton}, \citenamefont
  {Budd}, \citenamefont {Conrad}, \citenamefont {Johnson}, \citenamefont
  {King}, \citenamefont {Kinnel}, \citenamefont {Lamm}, \citenamefont
  {Lefmann}, \citenamefont {Marsh}, \citenamefont {McFarland}, \citenamefont
  {McNulty}, \citenamefont {Mishra}, \citenamefont {Naples}, \citenamefont
  {Quintas}, \citenamefont {Romosan}, \citenamefont {Sakumoto}, \citenamefont
  {Schellman}, \citenamefont {Sciulli}, \citenamefont {Seligman}, \citenamefont
  {Shaevitz}, \citenamefont {Smith}, \citenamefont {Spentzouris}, \citenamefont
  {Stern}, \citenamefont {Vakili}, \citenamefont {Yang},\ and\ \citenamefont
  {Yu}}]{Kim1998}%
  \BibitemOpen
  \bibfield  {author} {\bibinfo {author} {\bibfnamefont {J.~H.}\ \bibnamefont
  {Kim}}, \bibinfo {author} {\bibfnamefont {D.~A.}\ \bibnamefont {Harris}},
  \bibinfo {author} {\bibfnamefont {C.~G.}\ \bibnamefont {Arroyo}}, \bibinfo
  {author} {\bibfnamefont {L.}~\bibnamefont {de~Barbaro}}, \bibinfo {author}
  {\bibfnamefont {P.}~\bibnamefont {de~Barbaro}}, \bibinfo {author}
  {\bibfnamefont {A.~O.}\ \bibnamefont {Bazarko}}, \bibinfo {author}
  {\bibfnamefont {R.~H.}\ \bibnamefont {Bernstein}}, \bibinfo {author}
  {\bibfnamefont {A.}~\bibnamefont {Bodek}}, \bibinfo {author} {\bibfnamefont
  {T.}~\bibnamefont {Bolton}}, \bibinfo {author} {\bibfnamefont
  {H.}~\bibnamefont {Budd}}, \bibinfo {author} {\bibfnamefont {J.}~\bibnamefont
  {Conrad}}, \bibinfo {author} {\bibfnamefont {R.~A.}\ \bibnamefont {Johnson}},
  \bibinfo {author} {\bibfnamefont {B.~J.}\ \bibnamefont {King}}, \bibinfo
  {author} {\bibfnamefont {T.}~\bibnamefont {Kinnel}}, \bibinfo {author}
  {\bibfnamefont {M.~J.}\ \bibnamefont {Lamm}}, \bibinfo {author}
  {\bibfnamefont {W.~C.}\ \bibnamefont {Lefmann}}, \bibinfo {author}
  {\bibfnamefont {W.}~\bibnamefont {Marsh}}, \bibinfo {author} {\bibfnamefont
  {K.~S.}\ \bibnamefont {McFarland}}, \bibinfo {author} {\bibfnamefont
  {C.}~\bibnamefont {McNulty}}, \bibinfo {author} {\bibfnamefont {S.~R.}\
  \bibnamefont {Mishra}}, \bibinfo {author} {\bibfnamefont {D.}~\bibnamefont
  {Naples}}, \bibinfo {author} {\bibfnamefont {P.~Z.}\ \bibnamefont {Quintas}},
  \bibinfo {author} {\bibfnamefont {A.}~\bibnamefont {Romosan}}, \bibinfo
  {author} {\bibfnamefont {W.~K.}\ \bibnamefont {Sakumoto}}, \bibinfo {author}
  {\bibfnamefont {H.}~\bibnamefont {Schellman}}, \bibinfo {author}
  {\bibfnamefont {F.~J.}\ \bibnamefont {Sciulli}}, \bibinfo {author}
  {\bibfnamefont {W.~G.}\ \bibnamefont {Seligman}}, \bibinfo {author}
  {\bibfnamefont {M.~H.}\ \bibnamefont {Shaevitz}}, \bibinfo {author}
  {\bibfnamefont {W.~H.}\ \bibnamefont {Smith}}, \bibinfo {author}
  {\bibfnamefont {P.}~\bibnamefont {Spentzouris}}, \bibinfo {author}
  {\bibfnamefont {E.~G.}\ \bibnamefont {Stern}}, \bibinfo {author}
  {\bibfnamefont {M.}~\bibnamefont {Vakili}}, \bibinfo {author} {\bibfnamefont
  {U.~K.}\ \bibnamefont {Yang}}, \ and\ \bibinfo {author} {\bibfnamefont
  {J.}~\bibnamefont {Yu}},\ }\href {\doibase 10.1103/PhysRevLett.81.3595}
  {\bibfield  {journal} {\bibinfo  {journal} {Physical Review Letters}\
  }\textbf {\bibinfo {volume} {81}},\ \bibinfo {pages} {3595} (\bibinfo {year}
  {1998})}\BibitemShut {NoStop}%
\bibitem [{\citenamefont {Schienbein}\ \emph {et~al.}(2008)\citenamefont
  {Schienbein}, \citenamefont {Radescu}, \citenamefont {Zeller}, \citenamefont
  {Christy}, \citenamefont {Keppel}, \citenamefont {McFarland}, \citenamefont
  {Melnitchouk}, \citenamefont {Olness}, \citenamefont {Reno}, \citenamefont
  {Steffens},\ and\ \citenamefont {Yu}}]{Schienbein2008}%
  \BibitemOpen
  \bibfield  {author} {\bibinfo {author} {\bibfnamefont {I.}~\bibnamefont
  {Schienbein}}, \bibinfo {author} {\bibfnamefont {V.~A.}\ \bibnamefont
  {Radescu}}, \bibinfo {author} {\bibfnamefont {G.~P.}\ \bibnamefont {Zeller}},
  \bibinfo {author} {\bibfnamefont {M.~E.}\ \bibnamefont {Christy}}, \bibinfo
  {author} {\bibfnamefont {C.~E.}\ \bibnamefont {Keppel}}, \bibinfo {author}
  {\bibfnamefont {K.~S.}\ \bibnamefont {McFarland}}, \bibinfo {author}
  {\bibfnamefont {W.}~\bibnamefont {Melnitchouk}}, \bibinfo {author}
  {\bibfnamefont {F.~I.}\ \bibnamefont {Olness}}, \bibinfo {author}
  {\bibfnamefont {M.~H.}\ \bibnamefont {Reno}}, \bibinfo {author}
  {\bibfnamefont {F.}~\bibnamefont {Steffens}}, \ and\ \bibinfo {author}
  {\bibfnamefont {J.-Y.}\ \bibnamefont {Yu}},\ }\href {\doibase
  10.1088/0954-3899/35/5/053101} {\bibfield  {journal} {\bibinfo  {journal}
  {Journal of Physics G: Nuclear and Particle Physics}\ }\textbf {\bibinfo
  {volume} {35}},\ \bibinfo {pages} {053101} (\bibinfo {year}
  {2008})}\BibitemShut {NoStop}%
\bibitem [{\citenamefont {Bl{\"{u}}mlein}\ and\ \citenamefont
  {Tkabladze}(1999)}]{Blumlein1999}%
  \BibitemOpen
  \bibfield  {author} {\bibinfo {author} {\bibfnamefont {J.}~\bibnamefont
  {Bl{\"{u}}mlein}}\ and\ \bibinfo {author} {\bibfnamefont {A.}~\bibnamefont
  {Tkabladze}},\ }\href {\doibase 10.1088/0954-3899/25/7/350} {\bibfield
  {journal} {\bibinfo  {journal} {Journal of Physics G: Nuclear and Particle
  Physics}\ }\textbf {\bibinfo {volume} {25}},\ \bibinfo {pages} {1553}
  (\bibinfo {year} {1999})}\BibitemShut {NoStop}%
\bibitem [{\citenamefont {Seng}\ \emph {et~al.}(2019)\citenamefont {Seng},
  \citenamefont {Gorchtein},\ and\ \citenamefont {Ramsey-Musolf}}]{Seng2019b}%
  \BibitemOpen
  \bibfield  {author} {\bibinfo {author} {\bibfnamefont {C.~Y.}\ \bibnamefont
  {Seng}}, \bibinfo {author} {\bibfnamefont {M.}~\bibnamefont {Gorchtein}}, \
  and\ \bibinfo {author} {\bibfnamefont {M.~J.}\ \bibnamefont
  {Ramsey-Musolf}},\ }\href {\doibase 10.1103/PhysRevD.100.013001} {\bibfield
  {journal} {\bibinfo  {journal} {Physical Review D}\ }\textbf {\bibinfo
  {volume} {100}},\ \bibinfo {pages} {13001} (\bibinfo {year} {2019})},\
  \Eprint {http://arxiv.org/abs/1812.03352} {arXiv:1812.03352} \BibitemShut
  {NoStop}%
\bibitem [{\citenamefont {M{\"{a}}rkisch}\ \emph {et~al.}(2019)\citenamefont
  {M{\"{a}}rkisch}, \citenamefont {Mest}, \citenamefont {Saul}, \citenamefont
  {Wang}, \citenamefont {Abele}, \citenamefont {Dubbers}, \citenamefont
  {Klopf}, \citenamefont {Petoukhov}, \citenamefont {Roick}, \citenamefont
  {Soldner},\ and\ \citenamefont {Werder}}]{Markisch2019}%
  \BibitemOpen
  \bibfield  {author} {\bibinfo {author} {\bibfnamefont {B.}~\bibnamefont
  {M{\"{a}}rkisch}}, \bibinfo {author} {\bibfnamefont {H.}~\bibnamefont
  {Mest}}, \bibinfo {author} {\bibfnamefont {H.}~\bibnamefont {Saul}}, \bibinfo
  {author} {\bibfnamefont {X.}~\bibnamefont {Wang}}, \bibinfo {author}
  {\bibfnamefont {H.}~\bibnamefont {Abele}}, \bibinfo {author} {\bibfnamefont
  {D.}~\bibnamefont {Dubbers}}, \bibinfo {author} {\bibfnamefont
  {M.}~\bibnamefont {Klopf}}, \bibinfo {author} {\bibfnamefont
  {A.}~\bibnamefont {Petoukhov}}, \bibinfo {author} {\bibfnamefont
  {C.}~\bibnamefont {Roick}}, \bibinfo {author} {\bibfnamefont
  {T.}~\bibnamefont {Soldner}}, \ and\ \bibinfo {author} {\bibfnamefont
  {D.}~\bibnamefont {Werder}},\ }\href {\doibase
  10.1103/PhysRevLett.122.242501} {\bibfield  {journal} {\bibinfo  {journal}
  {Physical Review Letters}\ }\textbf {\bibinfo {volume} {122}},\ \bibinfo
  {pages} {242501} (\bibinfo {year} {2019})},\ \Eprint
  {http://arxiv.org/abs/1812.04666} {arXiv:1812.04666} \BibitemShut {NoStop}%
\bibitem [{\citenamefont {Walker-Loud}\ \emph {et~al.}(2020)\citenamefont
  {Walker-Loud}, \citenamefont {Berkowitz}, \citenamefont {Gambhir},
  \citenamefont {Brantley}, \citenamefont {Vranas}, \citenamefont {Bouchard},
  \citenamefont {Clark}, \citenamefont {Garron}, \citenamefont {Chang},
  \citenamefont {Joo}, \citenamefont {Kurth}, \citenamefont {Monge-Camacho},
  \citenamefont {Nicholson}, \citenamefont {Orginos}, \citenamefont {Monahan},\
  and\ \citenamefont {Rinaldi}}]{Walker-Loud2020}%
  \BibitemOpen
  \bibfield  {author} {\bibinfo {author} {\bibfnamefont {A.}~\bibnamefont
  {Walker-Loud}}, \bibinfo {author} {\bibfnamefont {E.}~\bibnamefont
  {Berkowitz}}, \bibinfo {author} {\bibfnamefont {A.~S.}\ \bibnamefont
  {Gambhir}}, \bibinfo {author} {\bibfnamefont {D.}~\bibnamefont {Brantley}},
  \bibinfo {author} {\bibfnamefont {P.}~\bibnamefont {Vranas}}, \bibinfo
  {author} {\bibfnamefont {C.}~\bibnamefont {Bouchard}}, \bibinfo {author}
  {\bibfnamefont {M.}~\bibnamefont {Clark}}, \bibinfo {author} {\bibfnamefont
  {N.}~\bibnamefont {Garron}}, \bibinfo {author} {\bibfnamefont {C.~C.}\
  \bibnamefont {Chang}}, \bibinfo {author} {\bibfnamefont {B.}~\bibnamefont
  {Joo}}, \bibinfo {author} {\bibfnamefont {T.}~\bibnamefont {Kurth}}, \bibinfo
  {author} {\bibfnamefont {H.}~\bibnamefont {Monge-Camacho}}, \bibinfo {author}
  {\bibfnamefont {A.}~\bibnamefont {Nicholson}}, \bibinfo {author}
  {\bibfnamefont {K.}~\bibnamefont {Orginos}}, \bibinfo {author} {\bibfnamefont
  {C.}~\bibnamefont {Monahan}}, \ and\ \bibinfo {author} {\bibfnamefont
  {E.}~\bibnamefont {Rinaldi}},\ }in\ \href {\doibase 10.22323/1.317.0020}
  {\emph {\bibinfo {booktitle} {Proceedings of The 9th International workshop
  on Chiral Dynamics - PoS(CD2018)}}},\ \bibinfo {series and number} {\bibinfo
  {number} {September}}\ (\bibinfo  {publisher} {Sissa Medialab},\ \bibinfo
  {address} {Trieste, Italy},\ \bibinfo {year} {2020})\ p.\ \bibinfo {pages}
  {020},\ \Eprint {http://arxiv.org/abs/1912.08321} {arXiv:1912.08321}
  \BibitemShut {NoStop}%
\bibitem [{\citenamefont {Falkowski}\ \emph {et~al.}(2017)\citenamefont
  {Falkowski}, \citenamefont {Gonz{\'{a}}lez-Alonso},\ and\ \citenamefont
  {Mimouni}}]{Falkowski2017}%
  \BibitemOpen
  \bibfield  {author} {\bibinfo {author} {\bibfnamefont {A.}~\bibnamefont
  {Falkowski}}, \bibinfo {author} {\bibfnamefont {M.}~\bibnamefont
  {Gonz{\'{a}}lez-Alonso}}, \ and\ \bibinfo {author} {\bibfnamefont
  {K.}~\bibnamefont {Mimouni}},\ }\href {\doibase 10.1007/JHEP08(2017)123}
  {\bibfield  {journal} {\bibinfo  {journal} {Journal of High Energy Physics}\
  }\textbf {\bibinfo {volume} {2017}},\ \bibinfo {pages} {123} (\bibinfo {year}
  {2017})}\BibitemShut {NoStop}%
\bibitem [{\citenamefont {Efrati}\ \emph {et~al.}(2015)\citenamefont {Efrati},
  \citenamefont {Falkowski},\ and\ \citenamefont {Soreq}}]{Efrati2015}%
  \BibitemOpen
  \bibfield  {author} {\bibinfo {author} {\bibfnamefont {A.}~\bibnamefont
  {Efrati}}, \bibinfo {author} {\bibfnamefont {A.}~\bibnamefont {Falkowski}}, \
  and\ \bibinfo {author} {\bibfnamefont {Y.}~\bibnamefont {Soreq}},\ }\href
  {\doibase 10.1007/JHEP07(2015)018} {\bibfield  {journal} {\bibinfo  {journal}
  {Journal of High Energy Physics}\ }\textbf {\bibinfo {volume} {2015}}
  (\bibinfo {year} {2015}),\ 10.1007/JHEP07(2015)018}\BibitemShut {NoStop}%
\bibitem [{\citenamefont {{The ATLAS and CMS
  collaborations}}(2016)}]{TheATLAScollaboration2016}%
  \BibitemOpen
  \bibfield  {author} {\bibinfo {author} {\bibnamefont {{The ATLAS and CMS
  collaborations}}},\ }\href {\doibase 10.1007/JHEP08(2016)045} {\bibfield
  {journal} {\bibinfo  {journal} {Journal of High Energy Physics}\ }\textbf
  {\bibinfo {volume} {2016}},\ \bibinfo {pages} {045} (\bibinfo {year}
  {2016})}\BibitemShut {NoStop}%
\bibitem [{\citenamefont {Severijns}\ \emph {et~al.}(2008)\citenamefont
  {Severijns}, \citenamefont {Tandecki}, \citenamefont {Phalet},\ and\
  \citenamefont {Towner}}]{Severijns2008}%
  \BibitemOpen
  \bibfield  {author} {\bibinfo {author} {\bibfnamefont {N.}~\bibnamefont
  {Severijns}}, \bibinfo {author} {\bibfnamefont {M.}~\bibnamefont {Tandecki}},
  \bibinfo {author} {\bibfnamefont {T.}~\bibnamefont {Phalet}}, \ and\ \bibinfo
  {author} {\bibfnamefont {I.~S.}\ \bibnamefont {Towner}},\ }\href {\doibase
  10.1103/PhysRevC.78.055501} {\bibfield  {journal} {\bibinfo  {journal}
  {Physical Review C}\ }\textbf {\bibinfo {volume} {78}},\ \bibinfo {pages}
  {055501} (\bibinfo {year} {2008})}\BibitemShut {NoStop}%
\bibitem [{\citenamefont {Naviliat-Cuncic}\ and\ \citenamefont
  {Severijns}(2009)}]{Naviliat-Cuncic2009}%
  \BibitemOpen
  \bibfield  {author} {\bibinfo {author} {\bibfnamefont {O.}~\bibnamefont
  {Naviliat-Cuncic}}\ and\ \bibinfo {author} {\bibfnamefont {N.}~\bibnamefont
  {Severijns}},\ }\href {\doibase 10.1103/PhysRevLett.102.142302} {\bibfield
  {journal} {\bibinfo  {journal} {Physical Review Letters}\ }\textbf {\bibinfo
  {volume} {102}},\ \bibinfo {pages} {142302} (\bibinfo {year}
  {2009})}\BibitemShut {NoStop}%
\bibitem [{\citenamefont {Towner}\ and\ \citenamefont
  {Hardy}(2015)}]{Towner2015}%
  \BibitemOpen
  \bibfield  {author} {\bibinfo {author} {\bibfnamefont {I.~S.}\ \bibnamefont
  {Towner}}\ and\ \bibinfo {author} {\bibfnamefont {J.~C.}\ \bibnamefont
  {Hardy}},\ }\href {\doibase 10.1103/PhysRevC.91.015501} {\bibfield  {journal}
  {\bibinfo  {journal} {Physical Review C}\ }\textbf {\bibinfo {volume} {91}},\
  \bibinfo {pages} {015501} (\bibinfo {year} {2015})}\BibitemShut {NoStop}%
\bibitem [{\citenamefont {Holstein}(1974)}]{Holstein1974}%
  \BibitemOpen
  \bibfield  {author} {\bibinfo {author} {\bibfnamefont {B.}~\bibnamefont
  {Holstein}},\ }\href
  {http://journals.aps.org/rmp/abstract/10.1103/RevModPhys.46.789} {\bibfield
  {journal} {\bibinfo  {journal} {Reviews of Modern Physics}\ }\textbf
  {\bibinfo {volume} {46}},\ \bibinfo {pages} {789} (\bibinfo {year}
  {1974})}\BibitemShut {NoStop}%
\bibitem [{\citenamefont {G{\"{o}}ckeler}\ \emph {et~al.}(1996)\citenamefont
  {G{\"{o}}ckeler}, \citenamefont {Horsley}, \citenamefont {Ilgenfritz},
  \citenamefont {Perlt}, \citenamefont {Rakow}, \citenamefont {Schierholz},\
  and\ \citenamefont {Schiller}}]{Gockeler1996}%
  \BibitemOpen
  \bibfield  {author} {\bibinfo {author} {\bibfnamefont {M.}~\bibnamefont
  {G{\"{o}}ckeler}}, \bibinfo {author} {\bibfnamefont {R.}~\bibnamefont
  {Horsley}}, \bibinfo {author} {\bibfnamefont {E.~M.}\ \bibnamefont
  {Ilgenfritz}}, \bibinfo {author} {\bibfnamefont {H.}~\bibnamefont {Perlt}},
  \bibinfo {author} {\bibfnamefont {P.}~\bibnamefont {Rakow}}, \bibinfo
  {author} {\bibfnamefont {G.}~\bibnamefont {Schierholz}}, \ and\ \bibinfo
  {author} {\bibfnamefont {A.}~\bibnamefont {Schiller}},\ }\href {\doibase
  10.1103/PhysRevD.53.2317} {\bibfield  {journal} {\bibinfo  {journal}
  {Physical Review D}\ }\textbf {\bibinfo {volume} {53}},\ \bibinfo {pages}
  {2317} (\bibinfo {year} {1996})}\BibitemShut {NoStop}%
\bibitem [{\citenamefont {Chambers}\ \emph {et~al.}(2017)\citenamefont
  {Chambers}, \citenamefont {Horsley}, \citenamefont {Nakamura}, \citenamefont
  {Perlt}, \citenamefont {Rakow}, \citenamefont {Schierholz}, \citenamefont
  {Schiller}, \citenamefont {Somfleth}, \citenamefont {Young},\ and\
  \citenamefont {Zanotti}}]{Chambers2017}%
  \BibitemOpen
  \bibfield  {author} {\bibinfo {author} {\bibfnamefont {A.~J.}\ \bibnamefont
  {Chambers}}, \bibinfo {author} {\bibfnamefont {R.}~\bibnamefont {Horsley}},
  \bibinfo {author} {\bibfnamefont {Y.}~\bibnamefont {Nakamura}}, \bibinfo
  {author} {\bibfnamefont {H.}~\bibnamefont {Perlt}}, \bibinfo {author}
  {\bibfnamefont {P.~E.~L.}\ \bibnamefont {Rakow}}, \bibinfo {author}
  {\bibfnamefont {G.}~\bibnamefont {Schierholz}}, \bibinfo {author}
  {\bibfnamefont {A.}~\bibnamefont {Schiller}}, \bibinfo {author}
  {\bibfnamefont {K.}~\bibnamefont {Somfleth}}, \bibinfo {author}
  {\bibfnamefont {R.~D.}\ \bibnamefont {Young}}, \ and\ \bibinfo {author}
  {\bibfnamefont {J.~M.}\ \bibnamefont {Zanotti}},\ }\href {\doibase
  10.1103/PhysRevLett.118.242001} {\bibfield  {journal} {\bibinfo  {journal}
  {Physical Review Letters}\ }\textbf {\bibinfo {volume} {118}},\ \bibinfo
  {pages} {242001} (\bibinfo {year} {2017})},\ \Eprint
  {http://arxiv.org/abs/1703.01153} {arXiv:1703.01153} \BibitemShut {NoStop}%
\end{thebibliography}%
\end{document}